\documentclass[a4paper,11pt]{scrartcl}
\usepackage[utf8x]{inputenc}			
\usepackage[T1]{fontenc}				
\usepackage{enumerate}
\usepackage[english]{babel}			
\usepackage{color}					
\usepackage{amsmath}					
\usepackage{amstext}					
\usepackage{amssymb}					
\usepackage{amsfonts}				
\usepackage{amsthm}					
\usepackage{mathrsfs}				
\usepackage{graphicx} 				
\usepackage{subfig}					
\usepackage{chemarrow}				
\usepackage{stackrel}				
\usepackage{cite}					
\usepackage{dsfont}					
\usepackage{tikz}
\usetikzlibrary{arrows, decorations.markings}
\usepackage{thmtools,thm-restate}

\numberwithin{equation}{section}
\allowdisplaybreaks[1]

\title{Fixation probabilities in populations under demographic fluctuations}
\author{Peter Czuppon, Arne Traulsen}
\date{\today}

\theoremstyle{definition}
\newtheorem{definition}{Definition}[section]

\theoremstyle{plain}
\newtheorem{theorem}[definition]{Theorem}
\newtheorem{lemma}[definition]{Lemma}
\newtheorem{corollary}[definition]{Corollary}

\theoremstyle{remark}
\newtheorem{remark}[definition]{Remark}

\begin{document}
\maketitle

\begin{abstract}
We study the fixation probability of a mutant type when introduced into a resident population. 
As opposed to the usual assumption of constant population size, we allow for stochastically varying population sizes. 
This is implemented by a stochastic competitive Lotka-Volterra model. 
The competition coefficients are interpreted in terms of inverse payoffs emerging from an evolutionary game. 
Since our study focuses on the impact of the competition values, we assume the same birth and death rates for both types. 
In this general framework, we derive an approximate formula for the fixation probability $\varphi$ of the mutant type under weak selection. 
The qualitative behavior of $\varphi$ when compared to the neutral scenario is governed by the invasion dynamics of an initially rare type. 
Higher payoffs when competing with the resident type yield higher values of $\varphi$. 
Additionally, we investigate the influence of the remaining parameters and find an explicit dependence of $\varphi$ on the mixed equilibrium value of the corresponding deterministic system (given that the parameter values allow for its existence).\\


\end{abstract}

\section{Introduction}
\label{intro}
The evolutionary dynamics of a mutant strain in a resident population is a well-studied topic in the field of population dynamics. 
Results concerning the fixation probability, the average fixation time or coexistence behavior can be applied in various biological fields, e.g. population genetics, bacterial evolution, viral dynamics or cancer initiation \cite{nowak:book:2006, Altrock:NRC:2015}. 
While the first theoretical analysis of such processes relied on deterministic differential equations, over the course of time more detailed models were studied describing the stochasticity of microscopic processes on the individual level. These individual based models can be approximated by the replicator equation (in the large population size limit) or be modeled by birth-death processes (in the case of two types) \cite{nowak:book:2006,sandholm:book:2010}. 
However, the dynamical evolution of the entire population is mostly neglected in these kinds of models and a constant population size is assumed instead. 
On the other hand, in population genetics and theoretical ecology, studies focused more on the effect that population dynamics have on the fixation probability rather than the concrete interaction mechanisms between the mutant and wild-type individuals \cite{ewens:Heredity:1967, kimura:PNAS:1974, otto:Genetics:1997}. 
More recently, researchers started investigating models connecting the stochastic interaction between individuals and stochastic population dynamics from a theoretical point of view \cite{lambert:AnAP:2005, lambert:TPB:2006, champagnat:AnAP:2007, parsons:TPB:2007a, parsons:TPB:2007b,melbinger:PRL:2010,cremer:PRE:2011,gabel:PRE:2013,constable:PNAS:2016}. 
For a historical overview on the calculation of fixation probabilities, see \cite{patwa:interface:2008}. 

To our knowledge, the first analytical approximation of fixation probabilities under stochastically varying population sizes is due to Lambert \cite{lambert:AnAP:2005, lambert:TPB:2006}. In these papers, the author analyzes models of interacting species by considering the corresponding diffusion equations under the constraint of weak selection. Going one step back on the descriptive scale and analyzing the Kolmogorov forward equation instead of its diffusion approximation, Champagnat and Lambert study the effect of various model parameters on the fixation probabilities and extend the previous results \cite{champagnat:AnAP:2007}. 
In parallel to these studies, Parsons and Quince examined the effect of variable growth rates on the fixation probability and mean fixation time in a two species system with stochastically varying population size \cite{parsons:TPB:2007a, parsons:TPB:2007b}. 
These results were later complemented and refined in \cite{parsons:genetics:2010}. 
Instead of focusing on variable growth rates, in this paper we concentrate on the effect of variable competition coefficients on the fixation probability.  

The model we will work with was introduced in \cite{huang:PNAS:2015}. 
It is a generalized two-type stochastic Lotka-Volterra-model which connects an evolutionary game with the competition coefficients of the model. 
Individuals of both species reproduce at constant rates and die 
spontaneously or based on competition within and between species. 
This leads to stochastically induced demographic fluctuations driven by interactions within the population. 
Our goal is to calculate the probability that a mutant takes over such a population of changing size. 

Recently, further models have been studied which connected game theoretical dynamics with exogenous population growth. 
For instance, Ashcroft et al.\ \cite{ashcroft:JTB:2017} consider a model with deterministic cell growth defined by a power law and stochastic species interactions derived from an evolutionary game. The authors rely on simulation results suggesting that the evolutionary outcome not only depends on the game played by the species, but also on the growth exponent of the power law governing the population growth.   
Constable et al.\ \cite{constable:PNAS:2016} study a public goods model. 
The authors analyze the invasion probability of producers and non-producers of the public good again under varying population sizes. 
Using a time-scale separation under a weak selection approximation, they find that producers can successfully invade a colony of non-producers even though they have a lower fitness than the resident type. 

The present paper is structured as follows: In Section \ref{sec:model} we describe the generalized Lotka-Volterra-model and restate some basic properties of the system, which were already described in \cite{huang:PNAS:2015}. 
In Section \ref{sec:fix_prob} we apply tools developed by Lambert \cite{lambert:TPB:2006} in order to derive a formula for the fixation probability in the weak selection limit. This allows us to interpret the impact of the competition coefficients separately. Furthermore, we compare the results for various competition matrices induced by different games with each other, i.e. the differences between coordination, coexistence and dominance games. Finally, in Section \ref{sec:single_mutant} we examine the fixation probability of a single mutant in a wild-type population, which allows us to compare our findings with those obtained in the previously studied settings, e.g. in finite but fixed population sizes. 

\section{Model}\label{sec:model}

The model we consider is a competitive Lotka-Volterra system consisting of two types, the mutant $X$ and the wild-type $Y$. We assume a well-mixed population, i.e. dynamics do not depend on the spatial arrangement of individuals, and a discrete state space describing the number of individuals of the two types, $X$ and~$Y$.\\

The evolution of the system is described by birth, death and competition processes, which we assume can be written in terms of chemical reactions. 
Each individual of the two types can reproduce or die independently of the other individuals. This leads to four reactions for the birth-death-processes,
\begin{equation} \label{eq:bd_dynamics} 
	X \xrightarrow{\beta_X} X + X, \qquad Y\xrightarrow{\beta_Y} Y+Y, \qquad X\xrightarrow{\gamma_X} \varnothing, \qquad Y\xrightarrow{\gamma_Y} \varnothing.
\end{equation}
Here, $\beta_X,\gamma_X$ and $\beta_Y,\gamma_Y$ denote the birth and death rates of the mutant and the wild-type, respectively. \\
Additionally, each individual competes with the other individuals and might die due to this process. These reactions occur at the rates
\begin{equation} \label{eq:competition_dynamics}
 X+X\xrightarrow{\frac{1}{aM}} X,\quad X+Y \xrightarrow{\frac{1}{bM}} Y,\quad X+Y\xrightarrow{\frac{1}{cM}} X,\quad Y+Y \xrightarrow{\frac{1}{dM}} Y,
\end{equation}
where $M$ controls the total population size in stationarity.

Later on, we interpret the competition rates as inverse payoffs of an evolutionary two-player game with payoff matrix 
\[ 
	\bordermatrix{ 
		& X & Y \cr
		X & a & b \cr
		Y & c & d
	}.
\]
This interpretation of the competition processes and a descriptive study of the stochastic competitive Lotka-Volterra system as well as a stability analysis of the stationary points of the corresponding deterministic system was performed by Huang et al.\ in \cite{huang:PNAS:2015}. 
This setup has the advantage that the average size of a monomorphic population reflects the payoffs. For example, a population of cooperators would be larger than a population of defectors, reflecting the fitness values within the population. 
The differential equations of the deterministic model read
\begin{equation}\label{eq:deterministic}
	\begin{aligned}
		\frac{dX}{dt} &= X\left(\beta_X-\gamma_X - \frac{X}{aM} - \frac{Y}{bM}\right),\\
		\frac{dY}{dt} &= Y\left(\beta_Y-\gamma_Y - \frac{X}{cM} - \frac{Y}{dM}\right).
	\end{aligned} 
\end{equation} 
For $a>c$ and $d>b$ as well as for $a<c$ and $d<b$, these equations have an internal stationary point where both species exist. It is given by
\[ (x^*,y^*) = \left(\frac{ac(b-d)}{bc-ad}(\beta_X-\gamma_X)M,\ \frac{bd(c-a)}{bc-ad}(\beta_Y-\gamma_Y)M\right). \]
Its stability depends on whether a coordination ($a>c$ and $b<d$) or a coexistence game ($a<c$ and $b>d$) is played. Additionally, we see that $M$ indeed characterizes the scale of the total population size. In the following, we will work with the fraction of mutants in the whole population given by $p=\frac{x}{x+y}$. We denote the steady state of this value by 
$$p^* = \frac{x^*}{x^* + y^*}=\frac{ac(b-d)}{ac(b-d)+bd(c-a)}.$$

Our goal is to extend the analysis of this particular system by approximating the fixation probability of the mutant type $X$ in a population of $Y$ individuals. 
The techniques we use rely on the theory of stochastic diffusions, see e.g.~\cite{ewens:book:2004}. 
Hence, we will work with the diffusion approximation of the above system; for a detailed derivation see Appendix~\ref{app:derivation_sde}. 
Letting $X(t)$ and $Y(t)$ be the number of mutant and wild-type individuals at time $t$, respectively, and setting $x(t)=\frac{X(t)}{M}$ and $y(t)=\frac{Y(t)}{M}$ we find
\begin{equation} \label{eq:diff_app}
	\begin{aligned}
		x(t) &= x(0) + \int_0^t x(s)\left((\beta_X-\gamma_X) - \frac{x(s)}{a} - \frac{y(s)}{b} \right) ds \\
			& \qquad \qquad \qquad + \frac{1}{\sqrt{M}}\int_0^t \sqrt{x(s)\left(\beta_X + \gamma_X + \frac{x(s)}{a} + \frac{y(s)}{b}\right)} dW^1(s),\\
		y(t) &= y(0) + \int_0^t y(s)\left((\beta_Y-\gamma_Y) - \frac{x(s)}{c} - \frac{y(s)}{d}\right) ds \\
			& \qquad \qquad \qquad + \frac{1}{\sqrt{M}} \int_0^t \sqrt{y(s)\left(\beta_Y + \gamma_Y + \frac{x(s)}{c} + \frac{y(s)}{d}\right)} dW^2(s),
	\end{aligned}
\end{equation}
where $W^1$ and $W^2$ are two independent, one-dimensional Brownian motions. The stochastic integrals are interpreted in the sense of It\^{o} \cite{kampen:book:1997,gardiner:book:2004}.

The solution of this system of differential equations is a two-dimensional Markov processes with infinitesimal generator given by (see Appendix \ref{app:derivation_sde} or \cite[Chapter 21]{kallenberg:book:2002})

\begin{equation}\label{eq:generator}
	\begin{aligned}	
		Gf(x,y) = 
		&\quad x\left(\beta_X-\gamma_X-\frac{x}{a}-\frac{y}{b}\right)\frac{\partial f}{\partial x} \\
		&+ y\left(\beta_Y-\gamma_Y-\frac{x}{c}-\frac{y}{d}\right) \frac{\partial f}{\partial y} \\
		&+ \frac{x}{2 M}\left(\beta_X + \gamma_X +\frac{x}{a}+\frac{y}{b}\right) \frac{\partial^2 f}{\partial x^2} \\
		&
		+ \frac{y}{2 M}\left(\beta_Y + \gamma_Y + \frac{x}{c} + \frac{y}{d}\right)\frac{\partial^2 f}{\partial y^2}.
	\end{aligned}
\end{equation}
We now proceed in deriving the fixation probability of the mutant type $X$.

\section{Fixation Probabilities}\label{sec:fix_prob}

The main result of this paper is the approximation of the probability of a mutant strain to fixate in a resident population of randomly fluctuating size under weak selection. Note first that due to the competition coefficients neither of the two species is able to go to~$\infty$ and hence each of them will die out at a (finite) random time \cite{lambert:TPB:2006}. We define fixation of the mutant $X$ as follows:
\begin{definition}[Fixation] 
Species $X$ fixates if for some $t\geq 0$ we have $y(t)=0$ and $x(t)>0$.
\end{definition}
In order to quantify the fixation probability we make use of the generator description of the model. Let $\varphi(x_0,y_0)$ be the fixation probability of species $X$ if the initial type-frequencies are $x_0$ and $y_0$. Then standard diffusion theory, see also \cite{ewens:book:2004, gardiner:book:2004} or Appendix \ref{app:fix_prob}, implies that $\varphi$ solves
\begin{equation}\label{eq:fix_pde}
	\left\{ \begin{array}{ll} G\varphi(x_0,y_0) = 0, & x_0,y_0 \geq 0,  \\ \varphi(x_0,0) = 1, & x_0>0, \\ \varphi(0,y_0) = 0, & y_0>0. 	\end{array}\right.
\end{equation}
In order to solve this partial differential equation we first do a parameter transformation to the coordinates $p=\frac{x}{x+y}$ and $z=x+y$, the fraction of $X$-individuals in the population and the whole population size, respectively. Given the same birth and death rates for both species, i.e. $\beta_X=\beta_Y=\beta$ and $\gamma_X=\gamma_Y=\gamma$, and noting that 
\[ \frac{1}{p^*}= 1+\frac{y^*}{x^*}=1 + \frac{bd}{ac}\cdot \frac{c-a}{b-d}, \]
the generator transforms to (the detailed calculations are given in Appendix~\ref{app:derivation_generator})
\begin{equation}\label{eq:whole_generator} \begin{aligned}
	\tilde{G}\varphi(p,z) &= \frac{p(1-p)}{d}\left(1-\frac{d}{b}\right)\left(1-\frac{p}{p^*}\right)\left(z+\frac{1}{M}\right) \frac{\partial \varphi}{\partial p}\\
			& + z\left[\beta-\gamma-\frac{z}{d}\left(1-p\left(2-\frac{d}{c}-\frac{d}{b}\right)+\left(1-\frac{d}{b}\right)\frac{p^2}{p^*}\right)\right] \frac{\partial \varphi}{\partial z}\\
			& + \frac{p(1-p)}{2zM}\left[\beta+\gamma + \frac{z}{d}\left(\frac{d}{b} + p\left(1+\frac{d}{a} - 2\frac{d}{b}\right) - \left(1-\frac{d}{b}\right)\frac{p^2}{p^*}\right)  \right] \frac{\partial^2 \varphi}{\partial p^2}\\
			& + \frac{p(1-p)z}{d M} \left(1-\frac{d}{b}\right)\left(\frac{p}{p^*} - 1\right) \frac{\partial^2 \varphi}{\partial p \partial z}\\
		& + \frac{z}{2M}\left[ \beta+\gamma + \frac{z}{d} \left(1 - p\left( 2- \frac{d}{c} -\frac{d}{b}\right) + \left(1-\frac{d}{b}\right)\frac{p^2}{p^*} \right) \right]\frac{\partial^2 \varphi}{\partial z^2}.
\end{aligned} 
\end{equation}
Equation \eqref{eq:fix_pde} translates to 
\begin{equation}\label{eq:fix_pde_transformed}
	\left\{ \begin{array}{ll} \tilde G\varphi(p_0,z_0) = 0, & p_0\in[0,1],z_0 \geq 0,  \\ \varphi(1,z_0) = 1, & z_0 > 0, \\ \varphi(0,z_0) = 0, & z_0 > 0. 	\end{array}\right.
\end{equation}
From now on, we drop the indices of $p_0$ and $z_0$ since the fixation probability always depends on the corresponding initial values.\\

Our goal is to approximate the solution of equation \eqref{eq:fix_pde_transformed}. 
Therefore, we start with the neutral setting which forms the basis of the subsequent calculations.

\subsection{Neutral model}
In formal terms, a neutral setting is given when individuals are exchangeable under labelling which in our case is equivalent to choosing a constant competition matrix, i.e.\ $a=b=c=d$. In this scenario the generator in equation \eqref{eq:whole_generator} simplifies to

\[ \begin{aligned}
	\tilde G \varphi_{neu}(p,z) &= z\left(\beta-\gamma -\frac{z}{a}\right) \frac{\partial \varphi_{neu}}{\partial z} + \frac{p(1-p)}{2zM}\left(\beta+\gamma+\frac{z}{a}\right)\frac{\partial^2 \varphi_{neu}}{\partial p^2}\\
	&\qquad \qquad \qquad \qquad \qquad \qquad \qquad \quad +  \frac{z}{2M}\left(\beta+\gamma+\frac{z}{a}\right)\frac{\partial^2 \varphi_{neu}}{\partial z^2}. 
\end{aligned} \]
Solving $\tilde G \varphi_{neu}(p,z) = 0$ with boundary conditions $$\varphi_{neu}(0,z) = 0 \text{ and } \varphi_{neu}(1,z) = 1 \text{ for } z>0$$ we obtain $\varphi_{neu}(p,z) = p$, the standard fixation probability of a mutant in an evolutionary process without selection. 

\subsection{Fixation probability under weak selection}
Based on the result of the neutral setting we approximate the fixation probability $\varphi(p,z)$ in the case of weak selection. 
In our model, this translates to the coefficients of the competition matrix being similar. 
To be more concrete we need the following conditions
\begin{enumerate}
	\item[(i)] $(1-\frac{d}{b})^2\ll 1$,
	\item[(ii)] $(1-\frac{d}{b})(2-\frac{d}{c}-\frac{d}{b}) \ll 1$,
	\item[(iii)] $(1-\frac{d}{b})(1+\frac{d}{a}-2\frac{d}{b}) \ll 1$ and
	\item[(iv)] $(1-\frac{d}{b})(\frac{1}{p^*}-2) \ll 1$. 
\end{enumerate}

In the following we will make use of asymptotic notation, i.e. 
\[ \begin{aligned}
	f(x)&= O(g(x)) \text{ for } x\to 0 \quad &\text{ iff } \qquad  &\lim_{x\to 0}\frac{f(x)}{g(x)}<\infty.
\end{aligned} \] 
We now state our main result.

\begin{restatable}[Fixation probability under weak selection]{theorem}{fixprob}
\label{thm:fix_prob}
	 Under conditions (i)-(iv) the solution of equation \eqref{eq:fix_pde_transformed} can be written as
	 \begin{equation}\label{eq:fix_prob} 
		\varphi(p,z) = p + p(1-p)\left(1-\frac{p}{p^*}\right)\left(1-\frac{d}{b}\right)\psi(z) + O\left(\left(1-\frac{d}{b}\right)^2\right),
	\end{equation}
	where $\psi(z)$ is independent of the initial frequency of mutants $p$ and solves
	\begin{equation}\label{eq:psi}
		\begin{aligned}
			0 &= \left(z+\frac{1}{M}\right) + z((\beta-\gamma)d-z) \psi'(z) - \frac{3}{zM}\left((\beta+\gamma)d+z\right) \psi(z)\\
			&\qquad \qquad \qquad \qquad \qquad \qquad \qquad \qquad \qquad \qquad + \frac{z}{2M}((\beta+\gamma)d+z) \psi''(z).
		\end{aligned}
	\end{equation}
\end{restatable}

\begin{remark}
	Note that this is basically a linearization around the neutral fixation probability and reduces to the neutral model if all payoff coefficients are equal, i.e. $\varphi(p,z)=p$ due to $1-\frac{d}{b}=0$. 
\end{remark}

The proof of the Theorem is given in Appendix \ref{sec:proof}. Basically, one applies the generator $\tilde G$ from equation \eqref{eq:whole_generator} 
to the formula stated in equation \eqref{eq:fix_prob}. Inserting conditions (i)-(iv) then gives the result. 

It seems remarkable that the initial population size does not affect the qualitative behaviour
of the fixation probability. However, the initial frequency of the mutant compared to the internal steady state and the payoffs $b$ and $d$ can change the sign of the first order effect under weak selection.
An interesting application is to consider fixation out of the neighbourhood of the internal steady state. 
Precisely at that point, we have for $\phi(p^*,z) = p^*$, as expected. 
For $p=p^*+\varepsilon$, we find
\begin{equation}\label{eq:fix_prob2} 
		\varphi(p^*+\varepsilon,z) = (p^*+\varepsilon)\left(1 - \frac{\varepsilon}{p^*}(1-p^*-\varepsilon)\left(1-\frac{d}{b}\right)\psi(z)\right) + O\left(\left(1-\frac{d}{b}\right)^2\right),
	\end{equation}
which implies that the fixation probability out of a neighborhood of a stable steady state $p^*$ ($d<b$) is smaller than neutral for positive deviations in $p$ and larger than neutral 
for negative deviations in $p$.
On the other hand, for an unstable steady state $p^*$ ($d>b$),
the fixation probability out of the neighbourhood is larger than neutral for positive deviations in $p$ and smaller than neutral for negative deviations in $p$. For a detailed study of fixation probabilities when leaving the deterministic steady state see also \cite{park:inprep:2017}. 

Next, we investigate different competition parameter constellations, i.e. conditions on the evolutionary game. We consider the following cases 
\begin{enumerate}
	\item[(a)]{\makebox[3cm][l]{coexistence game} \quad -- \quad $a<c$ and $b>d$;}
	\item[(b)]{\makebox[3cm][l]{coordination game} \quad -- \quad $a>c$ and $b<d$;}
	\item[(c)]{\makebox[3cm][l]{dominance game} \quad -- \quad $a>c, b>d$ or $a<c,b<d$.} 
\end{enumerate} 
The cases (a) and (b) allow for a mixed steady state in the deterministic model given in equation \eqref{eq:deterministic}. 
For coexistence games, this internal equilibrium is stable whereas for coordination games it is unstable, see for instance \cite{huang:PNAS:2015}. 

A qualitatively different picture arises in case (c). Here, either type $X$ or type $Y$ strictly dominates the other species in a game theoretic sense. This implies that the deterministic model only allows for single species equilibria where the stationary point of the dominant (inferior) type is stable (unstable). Thus, Theorem \ref{thm:fix_prob} does not hold in this case since $p^*$ does not tend to $\frac{1}{2}$. In fact, $p^*$ does not even exist. 
Instead we will replace condition~(iv) by an adapted version which then gives a similar approximation, see equation \eqref{eq:dom_approx}.

\subsection{Coexistence and Coordination Games}

In this section, we compare the resulting fixation probabilities in a coexistence and coordination game with the neutral fixation probability $\varphi_{neu}(p,z)=p$. In order to do so we need a Lemma characterizing the impact of the initial population size which we prove in Appendix \ref{app:psi_positive}.

\begin{lemma}
\label{lem:psi}
	The solution $\psi(z)$ of equation \eqref{eq:psi} is positive for all $z>0$.
\end{lemma}

\begin{figure}[ht]
	\centering
	\includegraphics[width=0.5\textwidth]{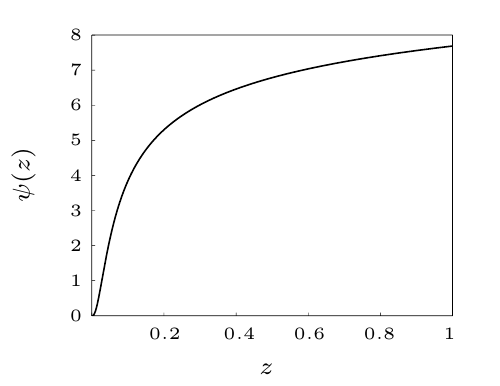}
    	\caption{The figure shows the numerical solution of equation~\eqref{eq:psi}. It remains positive and is growing with increasing $z$. Details on the numerical evalution of $\psi$ can be found in Appendix~\ref{app:numerics}. Parameters are given by $d=1, M=100, \beta=0.6, \gamma = 0.1$.}
    	\label{fig:psi_explicit}
\end{figure}

\begin{remark}
	In fact the function $\psi$ is a growing function in $z$ as can be seen in Figure~\ref{fig:psi_explicit}. This basically means that a larger initial population size affects the fixation probability stronger than an initially small population size where demographic effects are negligible.
\end{remark}

Now, we can state some immediate consequences of the fixation probability which follow from equation \eqref{eq:fix_prob}.

\begin{corollary}[Impact of competition parameters]\label{cor:comparison}
	Given the assumptions of Theorem~\ref{thm:fix_prob} we find the following:
	 \begin{enumerate} 
		\item For arbitrary $a,c>0$ and $p<p^*$ we have that $\varphi>\varphi_{neu}$ iff $b>d$.
		\item The probability of fixation is an increasing function in the competition parameter~$a$.
	\end{enumerate}
\end{corollary}

\begin{proof}
	Part 1.\ follows immediately by comparing $\varphi_{neu}$ with $\varphi$ from equation~\eqref{eq:fix_prob} and Lemma~\ref{lem:psi}. For part 2., we differentiate the representation of the fixation probability from equation~\eqref{eq:fix_prob} with respect to $a$ which gives
	\[ \frac{\partial \varphi}{\partial a} = p(1-p)\left(1-\frac{d}{b}\right) \psi(z) \frac{p}{(p^*)^2} \frac{bdc^2(b-d)}{(ac(b-d)+bd(c-a))^2} > 0. \]
	The last inequality holds for all choices of the parameter values which finishes the proof.
\end{proof}

\begin{remark}
	The first statement of the Corollary has the obvious implication that for a mutant to invade a resident population it is important to perform well against the wild-type. This also implies that species with lower single species equilibria (i.e. $a<d$) can have a higher chance of fixating than neutral. This can end up in an overall decrease of the overall population size. But still, as the second part of the Corollary shows, species with higher single-species equilibria also have a higher chance to fixate.
\end{remark}

Before turning to dominance games we take a brief look at some special cases in the context of coexistence and coordination games.

\paragraph{Symmetric and Asymmetric Games}
Here, we assume that $a=d$ and distinguish between symmetric games, i.e. $b=c$ and asymmetric games, $b\neq c$. In the case of symmetric games we have $\frac{a}{c}=\frac{d}{b}$ and thus $p^* = 1/2$. Therefore the fixation probability in equation \eqref{eq:fix_prob} simplifies to

\begin{equation}\label{eq:symmetric} 
	\varphi_{sym}(p,z) = p + p(1-p)(1-2p)\left(1-\frac{d}{b}\right)\psi(z). 
\end{equation}

\begin{figure}[ht]
	\centering
	\includegraphics[]{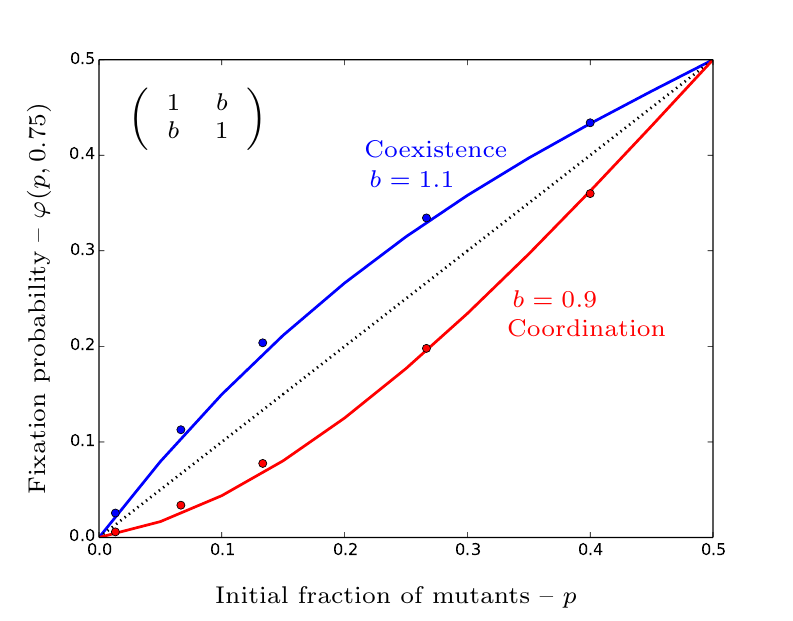}
    	\caption{The figure shows the fixation probability from equation \eqref{eq:symmetric} compared to the neutral fixation probability given by the dashed line. For coexistence games it is higher whereas in coordination games it is lower than the neutral values. The parameter are given by $a=d=1,b=c=0.9/1.1,M=100,\beta=0.6,\gamma=0.1$ and $z=0.75$. The bullets are averages taken over $100,000$ simulations. For the numerical solution of $\psi(z)$, see Appendix~\ref{app:numerics}.}
    	\label{fig:symmetric}
\end{figure}

Additionally, note that in this case we do not need assumption (iv) for the solution of the generator equation in \eqref{eq:generator_approx}. 
For an illustration of the fixation probability with some simulated data points, see Figure \ref{fig:symmetric}. 
For coexistence games ($b>d$) the fixation probability lies above the neutral line while for coordination games ($b<d$) the fixation probability is lower. 
Obviously, choosing $b$ closer to $d$ improves the analytical prediction due to the weak selection approximation in conditions (i)-(iii).

For asymmetric games, i.e. we still assume $a=d$ but now $b\neq c$, we obtain similar results. In this case, the fixation probability is given by

\begin{equation} \label{eq:asymmetric}
	\varphi_{asym}(p,z) = p + p(1-p)\left(1-\frac{p}{p^*}\right)\left(1-\frac{d}{b}\right)\psi(z).
\end{equation}

\begin{figure}[t!]
	\centering
	\includegraphics[]{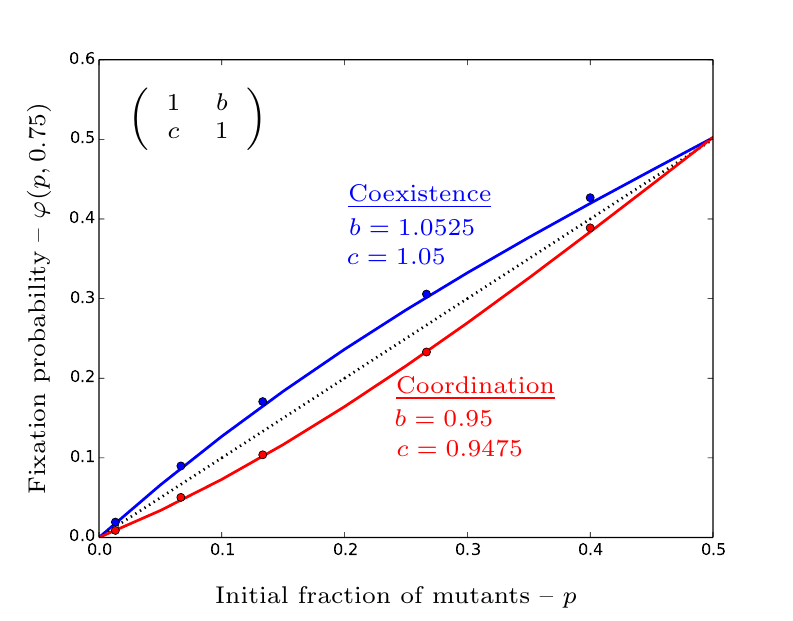}
	\caption{The fixation probability under asymmetric competition coefficients from equation \eqref{eq:asymmetric} is compared to the neutral fixation probability displayed by the dashed line. As in the symmetric case, coexistence games give higher and coordination games give lower values than the neutral model, respectively. Parameter are chosen as follows: $a=d=1,M=100,\beta=0.6,\gamma=0.1,z=0.75$ and $b,c$ as given in the figure. The points are averages taken over $100,000$ stochastic simulations of the original model and $\psi(z)$ is evaluated according to Appendix \ref{app:numerics}.}
	\label{fig:asymmetric}
\end{figure}

Dependent on whether $\frac{d}{b}>1$ (coordination) or $\frac{d}{b}<1$ (coexistence) the resulting fixation probability again lies below or above the neutral value, respectively, see also Figure \ref{fig:asymmetric}.

As already mentioned the condition for invasion, i.e. $\varphi_{asym}>\varphi_{neu}$ is $b>d$. This means that for a mutant to fixate in the resident population, it is primarily important to have a high payoff when playing against the resident, i.e. the more abundant type. However, when comparing $\varphi_{sym}$ and $\varphi_{asym}$ we see that here the parameter $c$ does play a role. To be more precise, whenever $b>c$ we have $\varphi_{asym}>\varphi_{sym}$. This condition resembles the shifting of the internal equilibrium of the deterministic system towards the mutant-axis, i.e. 
\[ p^*_{asym} > p^*_{sym}=\frac{1}{2} \qquad \text{iff} \qquad b>c.\]	

\subsection{Dominance Game}
In contrast to the coexistence and coordination game the dominance game does not have an internal equilibrium in its deterministic counterpart. This is due to one strategy strictly dominating the other. In terms of parameters this means that either $a>c$ and $b>d$ or $a<c$ and $b<d$ hold. As already mentioned, for the analysis of the fixation probability it does not make sense to assume condition (iv) which states that the internal equilibrium should be close to $\frac{1}{2}$. Hence, we can already infer that the analytical solution will not intersect with the neutral line due to one strategy being favored independently of its frequency. \\
For the calculation of $\varphi_{dom}$ we still assume conditions (i)-(iii) but instead of condition~(iv) we need the following:

\begin{enumerate}
	\item[(v)] $\left(1-\frac{d}{b}\right)\left(1+\frac{db}{ac}\frac{c-a}{b-d}\right)\ll 1$.
\end{enumerate}

This is plausible since the fraction in the last term should approximate $-1$ when considering a dominance game under weak selection, i.e. either $c>a$ and $d>b$ or $c<a$ and $d<b$. 

\begin{theorem}\label{thm:dominance}
	Under conditions (i)-(iii) and (v), we find
		\begin{equation}\label{eq:dom_approx}
			\varphi_{dom}(p,z) = p + p(1-p)\left(1-\frac{d}{b}\right)\psi(z) + O\left(\left(1-\frac{d}{b}\right)^2\right), 
		\end{equation}
	where $\psi(z)$ satisfies
	\begin{equation}\label{eq:psi_dom}
		\begin{aligned}
			0 &= \frac{1}{d}\left(z+\frac{1}{M}\right) + z\left(\beta-\gamma-\frac{z}{d}\right)\psi'(z)-\frac{1}{zM}\left(\beta+\gamma+\frac{z}{d}\right) \psi(z) \\
			& \qquad \qquad \qquad \qquad \qquad \qquad \qquad \qquad \qquad   + \frac{z}{2M}\left(\beta+\gamma+\frac{z}{d} \right) \psi''(z).
		\end{aligned} 
	\end{equation}
\end{theorem}

The proof is an imitation of the proof of Theorem \ref{thm:fix_prob} and therefore spared out.

We see that indeed $\varphi_{dom}$ is always larger or smaller than $\varphi_{neu}$ dependent on $b$ being larger or smaller than $d$, respectively. 
This finding is rather trivial, since we consider a dominance game and $b>d$ ensures the mutant being advantageous. 
More importantly, equation \eqref{eq:dom_approx} allows to calculate the
first order approximation of the neutral result. 

\section{Fixation of a single mutant}\label{sec:single_mutant}
\begin{figure}[ht]
	\centering
	\includegraphics[]{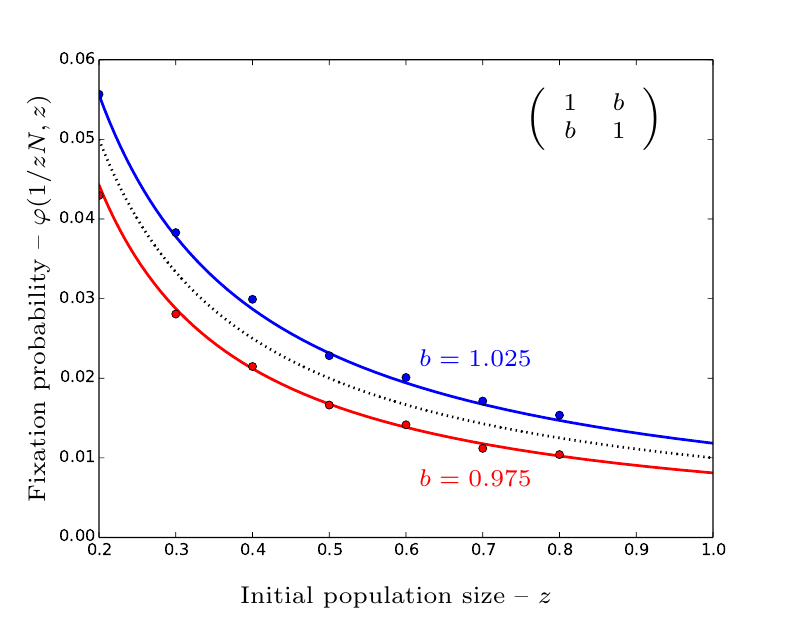}
	\caption{The fixation probability of a single mutant, equation \eqref{eq:symmetric}, under varying initial population sizes is shown in the case of a coexistence and coordination game. Again, these are separated by the neutral fixation probability (dashed line). The model parameters are set to $a=d=1,M=100,\beta=0.6,\gamma=0.1$ and $b=c$ as stated in the figure. The points are averages taken over $100,000$ simulations.}
	\label{fig:single_mutant}
\end{figure}

In this section, we consider the case that $p=\frac{1}{zM}$, i.e. initially there is exactly one mutant present in the population. This is probably the most realistic scenario as seen from a biologist's perspective. In contrast to the previous section we again focus on coexistence and coordination games, but now vary the initial population size instead of the number of mutants. 
As can be seen in Figure~\ref{fig:single_mutant} the probability of fixation is a decreasing function of the initial population sizes. This translates to the already observed fact that fixation of a mutant strain is more likely in a growing population than in a decreasing one, cf. \cite{kimura:PNAS:1974}. 
This can also be inferred from the formula describing the fixation probability since we are working in the weak selection limit, i.e. the governing part of $\varphi(p,z)$ is the initial frequency of mutants, here $\frac{1}{zM}$, which is decreasing for increasing $z$.

\begin{figure}[ht]
	\centering
	\includegraphics[]{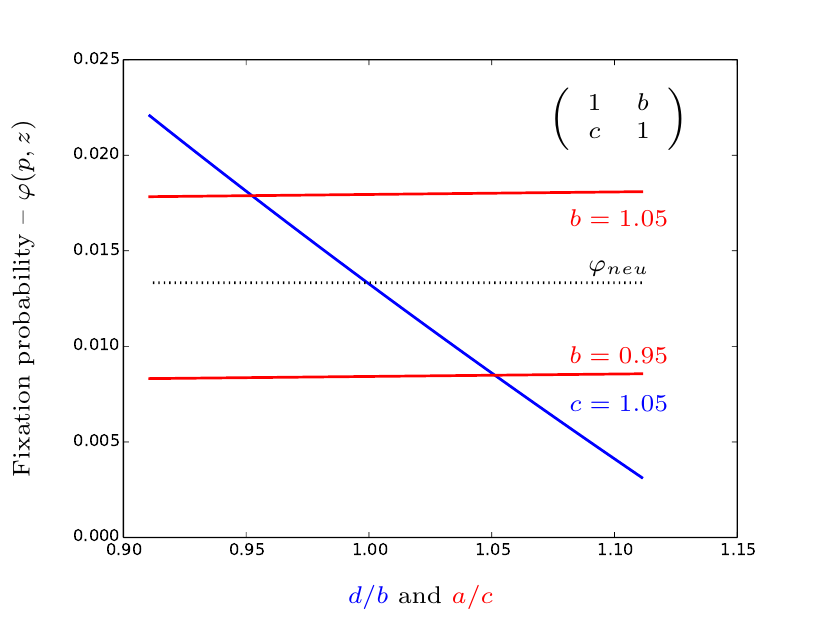}	
	\caption{This figure shows the impact of $b$ and $c$ on the fixation probability given in equation \eqref{eq:fix_prob}. Under variation of $c$ the fixation probability stays above or below the neutral fixation probability dependent on the choice of $b$. This is not true when we change $b$ but set $c$ to a constant value. This illustrates that indeed the evolutionary chance for a mutant to fixate is qualitatively independent of $c$ and completely determined by $b$ and $d$. Parameter values are: $a=d=1,z=0.75,M=100,p=\frac{1}{zM},\beta=0.6,\gamma=0.1$.}
	\label{fig:delta_alpha_figure}
\end{figure}

In models with constant population size $N$, the fixation probability of a mutant strain can be translated to the location of the mixed equilibrium of the corresponding replicator equation. 
More specifically, a mutant has a higher probability than neutral, i.e. $\frac{1}{N}$, to invade a resident population if the basin of attraction of the wild-type is smaller than $\frac{1}{3}$. This is referred to as the well-known $\frac{1}{3}$-law first derived in \cite{nowak:Nature:2004} and later generalized in scope in \cite{lessard:JMB:2007,lessard:DGAA:2011b}. \\
In the present implementation, a competitive Lotka-Volterra system, of a model with varying population size this simple rule does not hold anymore. Instead, the invasion probability only depends on the competition rates $b$ and $d$ describing the competition pressure of the species due to the resident type. However, this does not imply any properties of equilibria in the corresponding deterministic system. Even more, the choice of the parameters $a$ and $c$ does not affect the fixation probability when compared to the neutral case as was already pointed out in Corollary \ref{cor:comparison}. Thus, in the present model the difference between the neutral fixation probability and the probability of fixation under weak selection can be entirely described by the parameter $\frac{d}{b}$ and is independent of $\frac{a}{c}$ as can be seen in Figure \ref{fig:delta_alpha_figure}.\\
This kind of breakdown of the $\frac{1}{3}$-rule has also been observed by Ashcroft et al. in \cite{ashcroft:JTB:2017}, suggesting that a similar relation between fixation probability and the deterministic steady states does not exist in models with varying population size.

\section{Discussion and Conclusion}
The goal of this manuscript is the analysis of the invasion of a mutant strain when introduced into a wild-type population. 
Assuming a constant population size or deterministically varying population sizes this quantity has already been studied extensively. 
Here, we extend the analysis to systems including stochastic demographic fluctuations. 
Dealing with a competitive Lotka-Volterra model we are able to approximate the fixation probability under the weak selection assumption, i.e. the interaction rates between individuals just differ slightly. 
Therefore, we approximate the model in terms of stochastic diffusions and apply tools from stochastic diffusion theory to obtain an expression for the fixation probability.\\
We observe that the evolutionary success of a mutant mainly depends on its wild-type competition rate $b$. This is due to the resident type being more frequent initially yielding a higher probability for a mutant individual to interact with a resident type. This implies that a larger payoff for the mutant interacting with the wild-type ensures an enhanced fixation probability. This can be seen explicitly by the factor~$(1-\frac{d}{b})$ occurring in equation~\eqref{eq:fix_prob} which is the only term in the formula that can switch the sign given an initially rare mutant, i.e. $p<p^*$. \\
Still, the values $a$ and $c$ play a role in the overall evolutionary picture. While lowering~$c$ increases $p^*$ and thus the region where the invading type has a selective advantage, the parameter $a$ has an impact on the overall population size after fixation of the mutant strain. This might end in a decrease of the total number of individuals if $a<d$, even though the mutant has a selective advantage over the wild-type due to $b>d$. 

Furthermore, we studied the fixation probability of exactly one mutant in the initial population. Non-surprisingly and as already observed in systems with deterministic population growth/decrease, cf. \cite{kimura:PNAS:1974}, we see that the fixation probability monotonically decreases for increasing initial population sizes. Additionally, we find that in our system due to the varying population size the famous $\frac{1}{3}$-rule for fixed population size, see \cite{nowak:Nature:2004}, does not hold anymore. However, we can relate the deterministic equilibria to the intersection of the neutral fixation probability and its counterpart including selection, i.e. $\varphi_{neu}(p,z)=\varphi(p,z)$ if and only if $p\in\{0,1,p^*\}$.

The evolutionary result of populations under stochastically fluctuating population sizes has been studied in various scenarios over the last few years \cite{melbinger:PRL:2010, chotibut:PRE:2015, constable:PNAS:2016}. The stochasticity of the system as opposed to a deterministic modeling approach allows for different asymptotic behaviors and especially can reverse the deterministic behavior. This triggers the question for calculating fixation probabilities. 
We added some additional insight on the impact of the competition parameters on the fixation probability. 
Additionally, we showcase a method from stochastic diffusion theory and developed in \cite{lambert:TPB:2006} for approximating this quantity at least in the weak selection limit. Even though it is limited to the study of two interacting species, it is adaptable to many other models (and not only Lotka-Volterra-type systems) which include stochastic variation on the population size level. 


\bibliographystyle{alpha}
\bibliography{et.bib}   

\begin{appendix}

\section{Derivation of the diffusion approximation} \label{app:derivation_sde}
We derive the Focker-Planck equation corresponding to our model. 
The birth- and death-processes are given by
\begin{equation*} 
	X \xrightarrow{\beta_X} X + X, \qquad Y\xrightarrow{\beta_Y} Y+Y, \qquad X\xrightarrow{\gamma_X} \varnothing, \qquad Y\xrightarrow{\gamma_Y} \varnothing.
\end{equation*}
with $\beta_X,\gamma_X$ and $\beta_Y,\gamma_Y$ being the birth and death rates. 
The competition processes read

\begin{equation*} 
 X+X\xrightarrow{\frac{1}{aM}} X,\quad X+Y \xrightarrow{\frac{1}{bM}} Y,\quad X+Y\xrightarrow{\frac{1}{cM}} X,\quad Y+Y \xrightarrow{\frac{1}{dM}} Y,
\end{equation*}
with $M$ scaling the population size in stationarity. 
We follow the derivation of the Focker-Planck equation as done in \cite{huang:PNAS:2015} for the very same model. We set 
\[ \begin{aligned}
	T_X^+ &= \beta_x X, \qquad &T_X^- &= \gamma_x X + \frac{1}{aM}X(X-1) + \frac{1}{bM} XY,\\
	T_Y^+ &= \beta_y Y, \qquad &T_Y^- &= \gamma_y Y + \frac{1}{cM}XY + \frac{1}{dM}Y(Y-1)
\end{aligned} \]
as the transition rates of the system and calculate the infinitesimal generator $$(Gf)(X_t,Y_t)=\frac{\partial \mathbf{E} [f(X_t,Y_t)]}{\partial t}$$ of the process applied to a function $f(X_t,Y_t)$ dependent on the state $(X_t,Y_t)$ at time $t$ ($\mathbf{E}[\cdot]$ denotes the expectation of the stochastic system). Note that for $f(X_t,Y_t)=\mathds{1}_{\{X_t,Y_t\}}$ we retrieve the master equation. We obtain
\[\begin{aligned}
	(Gf)(X_t,Y_t)= \frac{\partial \mathbf{E} [f(X_t,Y_t)]}{\partial t}
		&= T_{X_t-1}^+ f(X_t-1,Y_t) + T_{X_t+1}^- f(X_t+1,Y_t) \\
		& \quad +T_{Y_t-1}^+ f(X_t,Y_t-1) + T_{Y_t+1}^- f(X_t,Y_t+1)  \\
		& \quad - (T_{X_t}^+ +T_{X_t}^- + T_{Y_t}^+ +T_{Y_t}^- ) f(X_t,Y_t).
\end{aligned} \]
Rescaling the parameters, i.e. setting $x=\frac{X}{M}$ and $y=\frac{Y}{M}$, we get

\[ \begin{aligned}
	\frac{\partial f(x_t,y_t)}{\partial t} 
	=& T_{X_t-1}^+ f(x_t-\tfrac{1}{M},y_t) + T_{X_t+1}^- f(x_t+\tfrac{1}{M},y_t) \\
	&+ T_{Y_t-1}^+ f(x_t,y_t-\tfrac{1}{M}) + T_{Y_t+1}^- f(x_t,y_t+\tfrac{1}{M})\\
	&- (T_{X_t}^+ + T_{X_t}^- + T_{Y_t}^+ + T_{Y_t}^-) f(x_t,y_t) \\
	=&  
		M \Big[ T_{x_t-\frac{1}{M}}^+ f(x_t-\tfrac{1}{M},y_t) + T_{x_t+\frac{1}{M}}^- f(x_t+\tfrac{1}{M},y_t) \\
		&+ T_{y_t-\frac{1}{M}}^+ f(x_t,y_t-\tfrac{1}{M}) + T_{y_t+\frac{1}{M}}^- f(x_t,y_t+\tfrac{1}{M})\\
		&- (T_{x_t}^+ + T_{x_t}^- + T_{y_t}^+ + T_{y_t}^-) f(x_t,y_t)\Big].
\end{aligned} \]

In the following we neglect the time subscript of the variables $x$ and $y$. Then, doing a Taylor expansion of the the function $f$ and the transition rates $T$ around $(x_t,y_t)$ up to the second order yields

\[ \begin{aligned}
	\frac{\partial f(x,y)}{\partial t} &= M\Bigg[ \left(T_{x}^+ - \frac{1}{M}\frac{\partial T_x^+}{\partial x} + \frac{1}{2M^2} \frac{\partial^2 T_x^+}{\partial x^2}\right)\left(f(x,y)- \frac{1}{M}\frac{\partial f(x,y)}{\partial x} + \frac{1}{2M^2} \frac{\partial^2 f(x,y)}{ \partial x^2}\right)\\
		&\quad  + \left(T_{x}^- + \frac{1}{M}\frac{\partial T_x^-}{\partial x} + \frac{1}{2M^2}\frac{\partial^2 T_x^-}{ \partial x^2}\right)\left(f(x,y)+ \frac{1}{M}\frac{\partial f(x,y)}{\partial x} + \frac{1}{2M^2} \frac{\partial^2 f(x,y)}{ \partial x^2}\right) \\
		&\quad  + \left(T_{y}^+ - \frac{1}{M} \frac{\partial T_y^+}{\partial y} + \frac{1}{2M^2} \frac{\partial^2 T_y^+}{ \partial y^2}\right)\left(f(x,y)- \frac{1}{M}\frac{\partial f(x,y)}{\partial y} + \frac{1}{2M^2} \frac{\partial^2 f(x,y)}{ \partial y^2}\right) \\
		&\quad + \left(T_{y}^- + \frac{1}{M} \frac{\partial T_y^-}{\partial y} + \frac{1}{2M^2} \frac{\partial^2 T_y^-}{ \partial y^2}\right)\left(f(x,y)+ \frac{1}{M} \frac{\partial f(x,y)}{\partial y} + \frac{1}{2M^2} \frac{\partial^2 f(x,y)}{ \partial y^2}\right) \\
		&\quad - (T_{x}^+ + T_x^- + T_y^+ + T_y^-) f(x,y)\Bigg]\\
		&= M\Bigg[ - \frac{1}{M} \left( \frac{\partial}{\partial x} \left(T_x^+\ f(x,y)\right)\right) + \frac{1}{2M^2}\left(\frac{\partial^2}{\partial x^2}\left( T_x^+\ f(x,y)\right)\right) + O\left(\frac{1}{M^3}\right) \\
		& \quad  + \frac{1}{M} \left( \frac{\partial}{\partial x} \left(T_x^-\ f(x,y)\right)\right) + \frac{1}{2M^2}\left(\frac{\partial^2}{\partial x^2}\left( T_x^-\ f(x,y)\right)\right) + O\left(\frac{1}{M^3}\right)\\
		&\quad - \frac{1}{M} \left( \frac{\partial}{\partial y} \left(T_y^+\ f(x,y)\right)\right) + \frac{1}{2M^2}\left(\frac{\partial^2}{\partial y^2}\left( T_y^+\ f(x,y)\right)\right) + O\left(\frac{1}{M^3}\right) \\
		&\quad + \frac{1}{N} \left( \frac{\partial}{\partial y} \left(T_y^-\ f(x,y)\right)\right) + \frac{1}{2M^2}\left(\frac{\partial^2}{\partial y^2}\left( T_y^-\ f(x,y)\right)\right) + O\left(\frac{1}{M^3}\right)\Bigg] \\
		&\approx -\frac{\partial}{\partial x}\left[(T_x^+ - T_x^-) f(x,y)\right] + \frac{1}{2M}\frac{\partial^2}{\partial x^2}\left[(T_x^+ + T_x^-) f(x,y)\right]\\
		&\quad -\frac{\partial}{\partial y}\left[(T_y^+ - T_y^-) f(x,y)\right] + \frac{1}{2M}\frac{\partial^2}{\partial y^2}\left[(T_y^+ + T_y^-) f(x,y)\right].
\end{aligned} \]

Inserting the terms for the transition rates $T_{x/y}^{+/-}$ gives the infinitesimal generator of the approximation which corresponds to the diffusion equation given in equation \eqref{eq:diff_app}, see for example \cite[Chapter 21]{kallenberg:book:2002}.

\section{Derivation of the fixation probability} \label{app:fix_prob}
Given the probability density $p_{(x_0,y_0)}(t;x,y)$ which describes the probability of the system given by the equations in \eqref{eq:diff_app} to be in state $(x,y)$ at time $t$ if started in $(x_0,y_0)$ the fixation probability is given by

\[ \varphi(t;x_0,y_0) = \int_0^t p_{(x_0,y_0)}(s;x>0,0)\, ds. \]
From \cite[Theorem 3.5]{lambert:AnAP:2005} we know that the population described by the dynamics in equation~\eqref{eq:diff_app} (or more generally a logistic Feller diffusion) goes extinct almost surely for times~$t$ large enough. Since then $p_{(x_0,y_0)}(t;0,0)\rightarrow 1$ for $t$ tending to infinity, the fixation probability $\varphi(x_0,y_0)=\lim_{t\rightarrow\infty} \varphi(t;x_0,y_0)$ satisfies 

\[ \frac{\partial \varphi(t;x_0,y_0)}{\partial t} \rightarrow \frac{\partial \varphi(x_0,y_0)}{\partial t} = 0, \qquad \text{ for } t\rightarrow\infty.\]
On the other hand we have
\[ \frac{\partial \varphi(x_0,y_0)}{\partial t} = \tilde G\varphi(x_0,y_0),\] 
where the operator $\tilde G$ is given in equation \eqref{eq:whole_generator}. Hence, we need to solve $\tilde G\varphi = 0$ with boundary conditions $\varphi(0,y) = 0$ and $\varphi(x,0)=1$ for $x,y>0$ which follow immediately from the model dynamics which then gives the partial differential equation with boundary conditions as stated in equation \eqref{eq:fix_pde}. 

\section{Derivation of transformed generator} \label{app:derivation_generator}
The generator of the system of stochastic differential equations given in equation \eqref{eq:diff_app} reads
\[ \begin{aligned} 
	Gf(x,y) &= x\left(\beta_X-\gamma_X-\frac{x}{a}-\frac{y}{b}\right)\frac{\partial f}{\partial x} + y\left(\beta_Y-\gamma_Y-\frac{x}{c}-\frac{y}{d}\right) \frac{\partial f}{\partial y} \\
		&\qquad \quad + \frac{x}{2 M}\left(\beta_X + \gamma_X +\frac{x}{a}+\frac{y}{b}\right) \frac{\partial^2 f}{\partial x^2} + \frac{y}{2 M}\left(\beta_Y + \gamma_Y + \frac{x}{c} + \frac{y}{d}\right)\frac{\partial^2 f}{\partial y^2}.
\end{aligned} \]
Doing a parameter transformation from the amount of individuals of each type $(x,y)$ to the fraction of mutants $p=\frac{x}{x+y}$ and the total population size $z=x+y$ we need to translate the derivatives into the new coordinate system. Now we have $x=pz$ and $y=(1-p)z$ which yields

\[ 
	\begin{aligned}
		\frac{\partial f}{\partial x}&=\frac{\partial f}{\partial p}\frac{\partial p}{\partial x} + \frac{\partial f}{\partial z}\frac{\partial z}{\partial x} = \frac{1-p}{z}\frac{\partial f}{\partial p} + \frac{\partial f}{\partial z},\\
		\frac{\partial f}{\partial y}&=\frac{\partial f}{\partial p}\frac{\partial p}{\partial x} + \frac{\partial f}{\partial z}\frac{\partial z}{\partial x}= -\frac{p}{z}\frac{\partial f}{\partial p} + \frac{\partial f}{\partial z},\\
		\frac{\partial^2 f}{\partial x^2} &= -\frac{2(1-p)}{z^2}\frac{\partial f}{\partial p} + \left(\frac{1-p}{z}\right)^2 \frac{\partial^2 f}{\partial p^2} + \frac{2(1-p)}{z}\frac{\partial^2 f}{\partial p \partial z} + \frac{\partial^2 f}{\partial z^2},\\
		\frac{\partial^2 f}{\partial y^2} &= \frac{2p}{z^2} \frac{\partial f}{\partial p} + \left(\frac{p}{z}\right)^2 \frac{\partial^2 f}{\partial p^2} - \frac{2p}{z}\frac{\partial^2 f}{\partial p \partial z} + \frac{\partial^2 f}{\partial z^2}.
	\end{aligned}
\]
Hence, the generator changes to 

\begin{align*}
		\tilde Gf(p,z)&= p(1-p)\left[\beta_X-\gamma_X -\beta_Y+\gamma_Y -\frac{pz}{a}-\frac{(1-p)z}{b} + \frac{pz}{c} + \frac{(1-p)z}{d}\right. \\
		&\quad \left. + \frac{1}{zM}\left(-\beta_X-\gamma_X+\beta_Y+\gamma_Y -\frac{pz}{a}-\frac{(1-p)z}{b} + \frac{pz}{c} + \frac{(1-p)z}{d}\right)\right] \frac{\partial f}{\partial p}\\
		&\quad + z\bigg[p (\beta_X-\gamma_X) + (1-p) (\beta_Y-\gamma_Y)  \\
		&\qquad \qquad \qquad \left. - \frac{p^2 z}{a}-\frac{p(1-p)z}{b} - \frac{p(1-p)z}{c}-\frac{(1-p)^2 z}{d}\right]\frac{\partial f}{\partial z}\\
		&\quad + \frac{p(1-p)}{2zM}\bigg[(\beta_X+\gamma_X)(1-p) + (\beta_Y+\gamma_Y) p  \\
		&\qquad \qquad \qquad \left. +\frac{p(1-p)z}{a}+\frac{(1-p)^2 z}{b} + \frac{p^2 z}{c} + \frac{p(1-p)z}{d}\right] \frac{\partial^2 f}{\partial p^2}\\
		&\quad + \frac{p(1-p)}{M}\left[\beta_X+\gamma_X - \beta_Y-\gamma_Y + \frac{pz}{a}+\frac{(1-p)z}{b}-\frac{pz}{c}-\frac{(1-p)z}{d}\right]\frac{\partial^2 f}{\partial p \partial z}\\
		&\quad + \frac{z}{2M}\bigg[ p (\beta_X+\gamma_X) + (1-p)(\beta_Y+\gamma_Y) \\
		&\qquad\qquad\qquad \left. +\frac{p^2 z}{a}+ \frac{p(1-p)z}{b} + \frac{p(1-p)z}{c}+ \frac{(1-p)^2 z}{d}\right]\frac{\partial^2 f}{\partial z^2}.
\end{align*}
Setting $\beta_X=\beta_Y=\beta$, $\gamma_X=\gamma_Y=\gamma$ and noting that $(p^*)^{-1} = 1+ \frac{bd}{ac}\frac{(c-a)}{(b-d)} $ we get

\[ \begin{aligned}
	\tilde Gf(p,z) &= p(1-p)\left(z+\frac{1}{M}\right) \left(\frac{1}{d}-\frac{1}{b}-p\left(\frac{1}{d}-\frac{1}{b}+\frac{1}{a}-\frac{1}{c}\right)\right)\frac{\partial f}{\partial p}\\
			& + z\left[\beta-\gamma-z\left(\frac{1}{d}-p\left(\frac{2}{d}-\frac{1}{c}-\frac{1}{b}\right)+p^2\left(\frac{1}{d}-\frac{1}{b}+\frac{1}{a}-\frac{1}{c}\right)\right)\right] \frac{\partial f}{\partial z}\\
			& + \frac{p(1-p)}{2zM}\left[\beta+\gamma + z\left(\frac{1}{b} + p\left(\frac{1}{d}+\frac{1}{a} - \frac{2}{b}\right)  -p^2\left(\frac{1}{d}-\frac{1}{b}+\frac{1}{a}-\frac{1}{c}\right) \right)  \right] \frac{\partial^2 f}{\partial p^2}\\
			&  + \frac{p(1-p)z}{M} \left(p\left(\frac{1}{d}-\frac{1}{b}+\frac{1}{a}-\frac{1}{c}\right) - \frac{1}{d}-\frac{1}{b}\right) \frac{\partial^2 f}{\partial p \partial z}\\
			& + \frac{z}{2M}\left[ \beta+\gamma + z\left(\frac{1}{d}-p\left(\frac{2}{d}-\frac{1}{c}-\frac{1}{b}\right)+p^2\left(\frac{1}{d}-\frac{1}{b}+\frac{1}{a}-\frac{1}{c}\right)\right) \right]\frac{\partial^2 f}{\partial z^2} \\
			= p&(1-p)\left(\frac{1}{d}-\frac{1}{b}\right)\left(z+\frac{1}{M}\right) \left(1-p\left(1+\frac{\frac{1}{a}-\frac{1}{c}}{\frac{1}{d}-\frac{1}{b}}\right)\right)\frac{\partial f}{\partial p}\\
			&  + z\left[\beta-\gamma-\frac{z}{d}\left(1-p\left(2-\frac{d}{c}-\frac{d}{b}\right)+p^2\left(1-\frac{d}{b}\right)\left(1+\frac{\frac{1}{a}-\frac{1}{c}}{\frac{1}{d}-\frac{1}{b}}\right)\right)\right] \frac{\partial f}{\partial z}\\
			& + \frac{p(1-p)}{2zM}\left[\beta+\gamma + \frac{z}{d}\left(\frac{d}{b} + p\left(1+\frac{d}{a} - 2\frac{d}{b}\right)  -p^2\left(1-\frac{d}{b}\right)\left(1+\frac{\frac{1}{a}-\frac{1}{c}}{\frac{1}{d}-\frac{1}{b}}\right) \right)  \right] \frac{\partial^2 f}{\partial p^2}\\
			& + \frac{p(1-p)z}{d M} \left(1-\frac{d}{b}\right)\left(p\left(1+\frac{\frac{1}{a}-\frac{1}{c}}{\frac{1}{d}-\frac{1}{b}}\right) - 1\right) \frac{\partial^2 f}{\partial p \partial z}\\
			& + \frac{z}{2M}\left[ \beta+\gamma + \frac{z}{d}\left(1-p\left(2-\frac{d}{c}-\frac{d}{b}\right)+p^2\left(1-\frac{d}{b}\right)\left(1+\frac{\frac{1}{a}-\frac{1}{c}}{\frac{1}{d}-\frac{1}{b}}\right)\right) \right]\frac{\partial^2 f}{\partial z^2} \\
			=p&(1-p)\frac{1}{d}\left(1-\frac{d}{b}\right)\left(z+\frac{1}{M}\right)\left(1-\frac{p}{p^*}\right) \frac{\partial f}{\partial p}\\
			& + z\left[\beta-\gamma-\frac{z}{d}\left(1-p\left(2-\frac{d}{c}-\frac{d}{b}\right)+\left(1-\frac{d}{b}\right)\frac{p^2}{p^*}\right)\right] \frac{\partial f}{\partial z}\\
			& + \frac{p(1-p)}{2zM}\left[\beta+\gamma + \frac{z}{d}\left(\frac{d}{b} + p\left(1+\frac{d}{a} - 2\frac{d}{b}\right) - \left(1-\frac{d}{b}\right)\frac{p^2}{p^*}\right)  \right] \frac{\partial^2 f}{\partial p^2}\\
			& + \frac{p(1-p)z}{d M} \left(1-\frac{d}{b}\right)\left(\frac{p}{p^*} - 1\right) \frac{\partial^2 f}{\partial p \partial z}\\
		& + \frac{z}{2M}\left[ \beta+\gamma + \frac{z}{d} \left(1 - p\left( 2- \frac{d}{c} -\frac{d}{b}\right)  + \left(1-\frac{d}{b}\right)\frac{p^2}{p^*} \right) \right]\frac{\partial^2 f}{\partial z^2},
\end{aligned} 
\]
which is precisely equation \eqref{eq:whole_generator}.

\section{Proof of Theorem \ref{thm:fix_prob}}\label{sec:proof}
Both, Theorem \ref{thm:fix_prob} and Theorem \ref{thm:dominance}, can be proved in a similar way. Due to this we will only give the proof of Theorem \ref{thm:fix_prob}. \\
First, we recall the the conditions needed for the Theorem:

\begin{enumerate}
	\item[(i)] $(1-\frac{d}{b})^2\ll 1$,
	\item[(ii)] $(1-\frac{d}{b})(2-\frac{d}{c}-\frac{d}{b}) \ll 1$,
	\item[(iii)] $(1-\frac{d}{b})(1+\frac{d}{a}-2\frac{d}{b}) \ll 1$ and
	\item[(iv)] $(1-\frac{d}{b})(\frac{1}{p^*}-2) \ll 1$. 
\end{enumerate}

The Theorem we want to prove is:

\fixprob*

\begin{proof}
	In order to determine the equation for $\psi(z)$ we apply the generator $\tilde G$ to $\varphi(p,z)$ in equation \ref{eq:generator}, which gives:

\[ \begin{aligned}
	\tilde G\varphi(p,z) &= \frac{p(1-p)}{d}\left(1-\frac{p}{p^*}\right)\left(1-\frac{d}{b}\right)\left(z+\frac{1}{M}\right)\left(1+O\left(1-\frac{d}{b}\right)\right) \\
		& \qquad + z\left[\beta-\gamma-\frac{z}{d}\left(1-p\left(2-\frac{d}{c}-\frac{d}{b}\right)+\left(1-\frac{d}{b}\right)\frac{p^2}{p^*}\right)\right] \times \\
		& \qquad \qquad \qquad \qquad \qquad \qquad \qquad \qquad \qquad p(1-p)\left(1-\frac{p}{p^*}\right)\left(1-\frac{d}{b}\right)\psi'(z)\\
		& \qquad + \frac{p(1-p)}{2zM}\left[\beta+\gamma + \frac{z}{d}\left(\frac{d}{b} + p\left(1+\frac{d}{a} - 2\frac{d}{b}\right) - \left(1-\frac{d}{b}\right)\frac{p}{p^*}\right)  \right]\times \\
		& \qquad \qquad \qquad \qquad \qquad \qquad \qquad \qquad \qquad \left[\frac{6p}{p^*} - 2 \left(1 + \frac{1}{p^*} \right) \right]\left(1-\frac{d}{b}\right)\psi(z)\\
		& \qquad + \frac{p(1-p)z}{d M} \left(1-\frac{d}{b}\right)\left(\frac{p}{p^*} - 1\right)O\left(1-\frac{d}{b}\right)\\
		& \qquad + \frac{z}{2M}\left[ \beta+\gamma + \frac{z}{d} \left(1 - p\left( 2- \frac{d}{c} -\frac{d}{b}\right) + \left(1-\frac{d}{b}\right)\frac{p}{p^*} \right) \right]\times \\
		& \qquad \qquad \qquad \qquad \qquad \qquad \qquad \qquad \qquad p(1-p)\left(1-\frac{p}{p^*}\right)\left(1-\frac{d}{b}\right)\psi''(z).
\end{aligned}\]
Next, applying the weak selection limit, i.e. using conditions (i)-(iii) we obtain

\begin{equation}\label{eq:generator_approx}
 \begin{aligned}
 	\tilde G\varphi(p,z) &\approx \frac{p(1-p)}{d}\left(1-\frac{p}{p^*}\right)\left(1-\frac{d}{b}\right)\left(z+\frac{1}{M}\right)\\
 		&\qquad \qquad  + z\left[\beta-\gamma-\frac{z}{d}\right]p(1-p)\left(1-\frac{p}{p^*}\right)\left(1-\frac{d}{b}\right)\psi'(z)\\
 		&\qquad \qquad + \frac{p(1-p)}{2zM}\left[\beta+\gamma + \frac{z}{b}\right]\left[\frac{6p}{p^*}-2 \left(1 +\frac{1}{p^*}\right) \right]\left(1-\frac{d}{b}\right)\psi(z) \\
 		& \qquad \qquad + \frac{z}{2M}\left[ \beta+\gamma + \frac{z}{d}\right]p(1-p)\left(1-\frac{p}{p^*}\right)\left(1-\frac{d}{b}\right)\psi''(z).
 \end{aligned}
\end{equation}
To simplify the term before $\psi(z)$, we observe
\[\begin{aligned} 	
	\beta+\gamma + \frac{z}{b} 
	= \beta+\gamma + \frac{z}{d}-\frac{z}{d} \left(1- \frac{d}{b} \right)
	= \beta+\gamma + \frac{z}{d}+O \left(1- \frac{d}{b} \right).	
\end{aligned} \]
and additionally
\[ 1+\frac{1}{p^*} = 1 + 2 + \frac{1}{p^*}-2=3 + \frac{1}{p^*}-2,\]
which in return yields
\[ \frac{6p}{p^*}-2 \left(1 +\frac{1}{p^*}\right) = 6\left(\frac{p}{p^*}-1\right) + \frac{1}{p^*}-2. \] 
Next, we insert these approximations into equation \eqref{eq:generator_approx} and apply condition (iv) such that the last term from above vanishes. Finally, setting $\tilde G\varphi=0$ and dividing by $\frac{p(1-p)}{d}(1-\frac{p}{p^*})(1-\frac{d}{b})$ gives

\[ 
	\begin{aligned}
		0 &= \left(z+\frac{1}{M}\right) + z((\beta-\gamma)d-z)\psi'(z) - \frac{3}{zM}\left((\beta+\gamma)d+z\right) \psi(z)  \\
		&\qquad \qquad \qquad \qquad \qquad \qquad \qquad \qquad \qquad \qquad + \frac{z}{2M}((\beta+\gamma)d+z) \psi''(z).
	\end{aligned}
\]
This yields equation \eqref{eq:psi} and finishes the proof.
\end{proof}

\section{Proof of Lemma \ref{lem:psi}}\label{app:psi_positive}
In this section, we prove that the solution $\psi(z)$ of equation \eqref{eq:psi}, i.e.\
\[ 
	\begin{aligned}
		0 &= \left(z+\frac{1}{M}\right) + z((\beta-\gamma)d-z)\psi'(z) - \frac{3}{zM}\left((\beta+\gamma)d+z\right) \psi(z)  \\
		&\qquad \qquad \qquad \qquad \qquad \qquad \qquad \qquad \qquad \qquad + \frac{z}{2M}((\beta+\gamma)d+z) \psi''(z).
	\end{aligned}
\]
is positive. Since the following section is very technical the reader who is satisfied with a numerical argument should skip this section and continue reading at Appendix \ref{app:numerics}.

\begin{remark}
The proof goes along the same lines as that of a similar result derived in \cite[Theorem 3.5]{lambert:TPB:2006}.
\end{remark}

Before we prove Lemma \ref{lem:psi} we recall and actually rewrite equation \eqref{eq:psi}. We know that $\psi(z)$ is the solution to the following ordinary differential equation:
\begin{equation}\label{eq:riccati}
	\begin{aligned}
		v(z)&= \left(z+\frac{1}{M}\right) \\
			&= - z((\beta-\gamma)d-z) \psi'(z) + \frac{3}{zM}\left((\beta+\gamma)d+z\right) \psi(z) - \frac{z}{2M}((\beta+\gamma)d+z) \psi''(z).
	\end{aligned}
\end{equation}
The first step is to rewrite equation \eqref{eq:riccati}. It is a Riccati-type equation and it is standard for these to do the following transformation: $h(z)=-\frac{\psi'(z)}{\psi(z)}$. This yields 
\[ h'(z) = -\frac{\psi''(z)}{\psi(z)} + h^2 (z) \]
and hence when considering the homogeneous differential equation \eqref{eq:riccati}, i.e.\ setting $v(z)=0$, we obtain
%
\begin{equation*}\label{eq:inhomogeneous}
	 z((\beta-\gamma)d -z)h + \frac{3}{zM}\left((\beta+\gamma)d + z\right) +\frac{z}{2M}((\beta+\gamma)d + z)(h'(z)-h^2(z))  = 0,
\end{equation*}
Rearranging terms, this equation reads
\begin{equation}\label{eq:homogeneous}
	h'(z)-h^2(z) + 2M\frac{(\beta-\gamma)d -z}{(\beta+\gamma)d + z}h(z) + \frac{6}{z^2}=0.
\end{equation}
Now, the proof of Lemma \ref{lem:psi} consist of the following two steps:
\begin{enumerate}
	\item[(i)] Solve equation \eqref{eq:homogeneous} and show that the solution is integrable in $[0,\infty)$. (Lemma~\ref{lem:homogeneous})
	\item[(ii)] Construct the solution of equation \eqref{eq:riccati} using the homogeneous solution characterized in Lemma~\ref{lem:homogeneous}.
\end{enumerate}	

Before we prove these two steps we state an auxiliary lemma, which we will make frequent use of.

\begin{lemma}\label{lem:integral}
	Let $f(t)$ be a real function with constant sign and rational behavior at $+\infty$. Then for $z\to\infty$ we have 
	\[\begin{aligned} \chi(z)&=6e^{2M(z-2d\beta\ln((\beta+\gamma)d+z)} \int_z^\infty f(t) e^{-2M(t -2d\beta\ln((\beta+\gamma)d+t))} dt\\
				&= 6 e^{m(z)}\int_z^\infty f(t)e^{-m(t)}dt \sim \frac{3f(z)}{M}.
	\end{aligned} \]
\end{lemma}
\begin{proof}
	The result follows by partial integration. See also Lemma A.1 in \cite{lambert:TPB:2006}. 
\end{proof}
We start with the first step. 

\begin{lemma}\label{lem:homogeneous}
	Equation \eqref{eq:homogeneous} has a unique and non-negative solution $h$ which satisfies
	\begin{enumerate}[(a)]
		\item $\limsup_{z\to\infty} z^2 h(z)\leq \frac{3}{M},$
		\item for $z\rightarrow 0+$ we find $h(z)= \frac{2}{z} + \frac{M(\beta-\gamma)}{\beta+\gamma} + O(z)$.	
	\end{enumerate}
\end{lemma}

\begin{remark}
Note, that the two statements characterizing the limit behavior of $h$ can be read off by forming the corresponding limits in equation \eqref{eq:homogeneous}.
\end{remark}

\begin{proof}
In order to show that $h$ is unique and non-negative, we set 
\[ \xi(z) = \int_0^z e^{m(x)}dx,\]
with $m(x):= 2M(x-2d\beta \ln((\beta+\gamma)d + x))$. Note that $\xi:[0,\infty) \rightarrow [0,\infty)$ is increasing and a bijection which allows us to define $\eta:=\xi^{-1}$. It holds $\eta'(x) = \exp(-m(\eta(x)))>0$. Furthermore, the derivative of $m$ satisfies 
\[ m'(x) = -2M\frac{(\beta-\gamma)d-x}{(\beta+\gamma)d+x}.\]
Next, let $w$ solve 
\begin{equation}\label{eq:w} 
	w' - w^2 = -6\left(\frac{\eta'}{\eta}\right)^2.
\end{equation}
Then, $h(z)=e^{m(z)}w(\xi(z))$ solves equation \eqref{eq:homogeneous} which can be seen by the following calculation:
\[ \begin{aligned}
	h'(z)-h^2(z) &= h(z)m'(z) + e^{m(z)} w'(\xi(z))\xi'(z) - e^{2m(z)}w^2(\xi(z)) \\
		&= h(z)m'(z) - e^{2m(z)} 6 \left(\frac{\eta'(\xi(z))}{\eta(\xi(z))}\right)^2 \\
		&= h(z)m'(z) - \frac{6}{z^2}.  
\end{aligned} \]

This means that $h$ solves equation \eqref{eq:homogeneous} if and only if $w$ solves equation \eqref{eq:w}. Redoing the proofs of Lemma 4.1, Lemma 4.2(i) and Lemma 4.3 of \cite{lambert:AnAP:2005} in our setting (which follow step-by-step in the same way and are therefore spared out) and arguing in the same vein as in the proof of Lemma 2.1 from \cite{lambert:AnAP:2005} (again step-by-step) we obtain that $w$ is the unique non-negative solution to equation \eqref{eq:w}. Furthermore, these results imply the following properties
\begin{enumerate}[(i)]
	\item $w\rightarrow 0$ for $z\rightarrow\infty$,
	\item $w\rightarrow \infty$ for $z\rightarrow 0$,
	\item $w$ decreases for $z$ tending to $\infty$,
	\item $w$ decreases in the neighborhood of $0+$,
	\item $w\leq \sqrt{6}\frac{\eta'}{\eta}$ in the neighborhood of $0+$ and $\infty$.
\end{enumerate}

Due to the definition of $h$ we can deduce that it is unique and non-negative and satisfies
\begin{enumerate}[(i)]
	\item $h\rightarrow \infty$ for $z\rightarrow 0+$,
	\item $0<h(z)<\frac{\sqrt{6}}{z}$.
\end{enumerate}

Next we examine the limit behavior of $h$. Mimicking the proof of Lemma 3.4 in \cite{lambert:TPB:2006} we show 
\[ \limsup_{z\rightarrow\infty} h(z) z^2 \leq \frac{1}{2M}. \]
In order to prove this, we set 
\[	\chi(z)=6\exp\left(m(z)\right) \int_z^\infty x^{-2} \exp\left(-m(x))\right) dx, \]
with 
\[ \chi'(z)  = m'(z)\chi(z)-\frac{6}{z^2} = -2M\frac{(\beta-\gamma)d-z}{(\beta+\gamma)d +z} \chi(z) -\frac{6}{z^2}.\]
This yields
\[\begin{aligned}
	(h-\chi)'(z) & = h(z)m'(z) + e^{2m(z)} w'(\xi(z)) - m'(z)\chi(z) + \frac{6}{z^2}\\
		&= h(z)m'(z) - m'(z)\chi(z) + e^{2m(z)} w^2(\xi(z))\\
		&> m'(z)(h(z)-\chi(z)).
\end{aligned} \]
For $z$ large enough, we have that $m'(z)>0$. 
Hence, whenever $h(z)>\chi(z)$ this gives $(h-\chi)'(z)>0$ which means that from that point on $h>\chi$. 
By Gronwall-type reasoning (see \cite[Lemma 3.4]{lambert:TPB:2006} and for the Lemma of Gronwall \cite[Theorem A.5.1]{ethier:book:1986}) we see that then $h$ tends to infinity. 
However, this is a contradiction to $h(z)\leq \frac{\sqrt{6}}{z}$. Thus, $h<\chi$ and therefore with Lemma~\ref{lem:integral} we have $\limsup_{z\rightarrow\infty} h(z)z^2 \leq \frac{3}{M}$. This shows statement~(a).\\
Next we turn our attention to the limit behavior of $h$ when $z$ approaches $0$ from above. Therefore, instead of $z>0$ we consider $c:=-z$ which simplifies the following reasoning. We define
\[ \zeta(c):= -\frac{c h(-c) + 2}{c^3} e^{-m(-c)}.\]
Noting that 
\[ h'(-c) +\frac{6}{c^2} - h(-c) m'(-c) = h^2(-c) \]
we get
\[\begin{aligned}
	\zeta'(c) &= -\frac{(h(-c)-c h'(-c))c^3 - 3c^2 (c h(-c)+2)}{c^6}e^{-m(-c)} -\frac{c h(-c)+2}{c^3}m'(-c)e^{-m(-c))} \\
		&= -\frac{1}{c^3}\left(-2h(-c) -c\left(h'(-c)+\frac{6}{c^2}-m'(-c)h(-c)\right) + 2m'(-c)\right)e^{-m(-c)}\\
		&= -\frac{1}{c^3}\left(-h(-c)(2+ch(-c))\right)e^{-m(-c)}-\frac{2}{c^3}e^{-m(-c)} \\
		&=-h(-c)\zeta(c) - \frac{2}{c^3}m'(-c) e^{-m(-c))}\\
		&=-h(-c)\zeta(c) + \frac{4M}{c^3}\frac{(\beta-\gamma)d+c}{(\beta+\gamma)d -c}e^{-m(-c)}.
\end{aligned} \]
For $c<0$ close enough to $0$ we see that $\zeta$ has constant sign in the neighborhood of $0-$ since if for some $c_0$ we have $\zeta(c_0)=0$ then $\zeta'(c_0)<0$. Thus, we have either (1) $\zeta>0$ or (2) $\zeta<0$ in the neighborhood of $0-$. This yields in case (1) 
\[ |\zeta(c)|' = \zeta'(c) <  \frac{2}{|c|^3}|m'(-c)| e^{-m(-c)} \]
and in case (2)
\[ |\zeta(c)|' = -\zeta'(c) = h(-c)\zeta(c) + \frac{2}{c^3} m'(-c)e^{-m(-c)} < \frac{2}{|c^3|}|m'(-c)| e^{-m(-c)}.\]
Thus, in both cases $|c^3| |\zeta(c)|'$ is strictly bounded from above near $0-$ which yields that $|\zeta(c)|\in O(\frac{1}{c^2})$. But this means that $h(z)\sim \frac{2}{z}$ for $z\rightarrow 0+$. \\
For the second order term we consider the auxiliary function
\[ \tau(z) := \frac{h(z)-\frac{2}{z}}{z^4}.\]
Again, we calculate the derivative and obtain
\[ \begin{aligned}
	\tau'(z) &= \frac{(h'(z)+\frac{2}{z^2})z^4 -4z^3 h(z)+8z^2}{z^8}\\
		&= \frac{1}{z^4}\left(h'(z)-h^2(z) + h^2(z) + \frac{2}{z^2} -4\frac{h(z)}{z} + \frac{8}{z^2}\right)\\
		&= \frac{1}{z^4}\left(h(z)m'(z) -\frac{6}{z^2} + \frac{10}{z^2} + h^2(z) - 4\frac{h(z)}{z}\right)\\
		&= \frac{1}{z^4}\left( \left(h(z)-\frac{2}{z}\right)^2 + m'(z)\left(h(z)-\frac{2}{z}\right) + m'(z)\frac{2}{z}\right)\\
		&=\frac{1}{z^4}\left(h(z)-\frac{2}{z}\right)^2 + \frac{m'(z)}{z^4}\left(h(z)-\frac{2}{z}\right) +\frac{1}{z^4}\left( \frac{4M}{(\beta+\gamma)d + z} - \frac{4M(\beta-\gamma)d}{z((\beta+\gamma)d+z)}\right).
\end{aligned} \]
Multiplying with $z^4$ we see that the first three terms on the right side are bounded for $z\rightarrow 0+$ whereas the last term is of order $z^{-1}$. This yields
\[ \tau(z) \approx \frac{1}{z^4}\frac{M(\beta-\gamma)}{(\beta+\gamma)} + O\left(\frac{1}{z^3}\right) \qquad \text{for } z\rightarrow 0+.\]
This finishes the proof of statement (b), i.e. $h(z) = \frac{2}{z} + \frac{M(\beta-\gamma)}{\beta+\gamma} + O(z)$ for $z\to 0+$. 
\end{proof}

Based on these estimates of the homogeneous solution $h$ we can now continue by solving the inhomogeneous equation \eqref{eq:psi}, see also equation \eqref{eq:riccati}. Our goal now is to prove Lemma \ref{lem:psi} from the main text which states that the function $\psi(z)$ solving this equation is positive.

\begin{proof}[Proof of Lemma \ref{lem:psi}]
Again, we follow the reasoning of the corresponding proof given in \cite{lambert:TPB:2006}.

First of all note that due to Lemma \ref{lem:homogeneous} (a) and (b) $h$ is integrable at $\infty$ and that as $z\rightarrow 0+$ we find

\begin{equation}\label{eq:hom_int}
	\int_z^\infty h(t)dt = \int_z^1 h(t)-\frac{2}{t}dt + \int_z^1 \frac{2}{t}dt + \int_1^\infty h(t) dt = \alpha + \ln(z^{-2}) + O(z),
\end{equation}
where $\alpha$ is given by
\[ \alpha:= \int_z^1 h(t)-\frac{2}{t}dt + \int_1^\infty h(t) dt.\]
Therefore we can define 
\begin{equation}\label{eq:psi_new_def}
	B(z):= \psi(z) \exp\left(-\int_z^\infty h(t) dt\right), 
\end{equation}
where $\psi$ is the solution of equation \eqref{eq:riccati} and h solves equation \eqref{eq:homogeneous}. In the following we will prove a representation of $B(z) = B(0) + \int_0^z B'(x)dx$ which will then give an explicit expression for $\psi(z)$. This expression will then show that $\psi$ is indeed positive for all $z$. \\
So, let us start by analyzing $B(z)$. 
Differentiating $B(z)$ gives
\[ \begin{aligned}
	B'(z) &= \left( \psi'(z) + \psi(z)h(z)\right)\exp\left(-\int_z^\infty h(t) dt\right),\\
	B''(z) &= \left( \psi''(z) + 2\psi'(z)h(z)+ \psi(z)h'(z)  + \psi(z)h^2(z) \right)\exp\left(-\int_z^\infty h(t) dt\right).
\end{aligned} \] 
Next, we look for a combination of these derivatives such that these give the inhomogeneous solution $v(z) = z+\frac{1}{M}$. This will then allow us to write down an explicit expression of $B'(z)$ which can then be translated to an explicit expression of $\psi$. Using that $\psi$ and $h$ are solutions of the differential equations given in \eqref{eq:riccati} and \eqref{eq:homogeneous}, respectively, we calculate
\begin{equation}\label{eq:aux_a}
	\begin{aligned}
		&\frac{1}{6}B''(z) - \frac{1}{3}h(z)B'(z) + \frac{M}{3}\frac{(\beta-\gamma)d-z}{(\beta+\gamma)d+z}B'(z)\\
		&\quad = \frac{1}{6} \left( \psi''(z) + 2\psi'(z)h(z)+ \psi(z)h'(z)  + \psi(z)h^2(z) \right)e^{-\int_z^\infty h(t) dt} \\
		& \quad \quad - \frac{1}{3}h(z)\left( \psi'(z) + \psi(z)h(z)\right)e^{-\int_z^\infty h(t) dt} \\
		& \quad \quad \quad + \frac{M}{3}\frac{(\beta-\gamma)d-z}{(\beta+\gamma)d+z}\left( \psi'(z) + \psi(z)h(z)\right)e^{-\int_z^\infty h(t) dt}\\
		& \quad = \left(\frac{1}{6}\psi''(z) + \frac{M}{3}\frac{(\beta-\gamma)d-z}{(\beta+\gamma)d+z}\psi'(z) + \frac{1}{3}\psi'(z)h(z)-\frac{1}{3}\psi'(z)h(z)\right.\\
		& \quad \quad +  \psi(z)\underbrace{\left(\frac{1}{6}(h'(z)+h^2(z))-\frac{1}{3}h^2(z) + \frac{M}{3}\frac{(\beta-\gamma)d-z}{(\beta+\gamma)d+z} h(z)\right)}_{=-\frac{1}{z^2} \text{ cf. \eqref{eq:homogeneous}}}\Bigg)e^{-\int_z^\infty h(t)dt}\\
		& \quad = \left(\frac{1}{6}\psi''(z) + \frac{M}{3}\frac{(\beta-\gamma)d-z}{(\beta+\gamma)d+z} \psi'(z) -\frac{1}{z^2}\psi(z)\right)e^{-\int_z^\infty h(t)dt}\\
		&\quad \stackrel{\eqref{eq:riccati}}{=} -v(z)\frac{M}{3z((\beta+\gamma)d+z)}e^{-\int_z^\infty h(t)dt}.
	\end{aligned}
\end{equation}

In order to determine $B(z)$ we make use of another auxiliary function
\begin{equation}\label{eq:aux_c}
	C(z):= \frac{1}{6}B'(z)\exp\left(2\int_z^\infty h(t) dt - m(z)\right).
\end{equation}
Again we calculate the derivative and, this time applying equation \eqref{eq:aux_a}, we obtain
\[\begin{aligned}
	C'(z) &= \left(\frac{1}{6}B''(z) -\frac{1}{3}h(z)B'(z) + \frac{M}{3}\frac{(\beta-\gamma)d-z}{(\beta+\gamma)d+z} B'(z)\right) e^{\int_z^\infty h(t)dt - m(z)} \\
		&=-\frac{M v(z)}{3z((\beta+\gamma)d+z)}e^{\int_z^\infty h(t)dt - m(z)}. 
\end{aligned} \]
Integrating from $z$ to $\infty$ gives
\[\begin{aligned}
	C(z) = C(\infty) + \int_z^\infty \frac{M v(t)}{3t((\beta+\gamma)d+t)}e^{\int_t^\infty h(s)ds - m(t)}dt,
\end{aligned} \]
which with equation \eqref{eq:aux_c} yields

\begin{equation}\label{eq:aux_b}
	\begin{aligned}
		B'(z) &= 6 C(\infty) e^{-2\int_z^\infty h(t) dt + m(z)} \\
		&\qquad + 6 e^{-2\int_z^\infty h(t) dt + m(z)} \int_z^\infty \frac{M v(t)}{3t((\beta+\gamma)d+t)}e^{\int_t^\infty h(s)ds -m(t)}dt.
	\end{aligned}
\end{equation}

Applying Lemma \ref{lem:integral} we see that the second term behaves like 
$$\frac{v(z)}{z((\beta+\gamma)d+z)}=\frac{z+\frac{1}{M}}{z((\beta+\gamma)d+z)}, \text{ as } z\to\infty,$$ 
which implies that 
\[ B'(z)\sim 6 B(\infty) e^{m(z)}, \]
as $z\to \infty$. Due to equation \eqref{eq:psi_new_def} this means that $\psi$ increases with $z$ faster than exponential which we show is not true. Therefore, $C(\infty)$ needs to be zero giving an explicit expression for $B'$. \\
Hence, let us consider the model for very large values of $z$. This implies that the system is governed by the quadratic competition terms and can be approximated by the corresponding ODE-system which reads:
\[\begin{aligned}
	\dot{x}_t &= - \frac{x_t^2}{a}-\frac{x_t y_t}{b},\\
	\dot{y}_t &= - \frac{x_t y_t}{c} -\frac{y_t^2}{d}.
\end{aligned} \]
The dynamics of the fraction of mutants, i.e. $p$, is then given by
\[ \begin{aligned}
	\dot{p}_t &= \frac{dx_t(x_t+y_t) -x_t(dx_t +dy_t)}{(x_t+y_t)^2} \\
		&= -\frac{p^2 z}{a} - \frac{p(1-p)z}{b} + \frac{pz}{z^2} \left(\frac{p^2z^2}{a} + \frac{p(1-p)z^2}{b}+\frac{p(1-p)z^2}{c}+\frac{(1-p)^2 z^2}{d}\right)\\
		&= zp(1-p)\left(\frac{(1-p)}{d}+\frac{p}{c}-\frac{(1-p)}{b}-\frac{p}{a}\right)\\
		&= zp(1-p)\left(\frac{1}{d}-\frac{1}{b}\right)\left(1-p\left(1+\frac{\frac{1}{a}-\frac{1}{c}}{\frac{1}{d}-\frac{1}{b}}\right) \right)\\
		&= \frac{zp(1-p)}{d}\left(1-\frac{d}{b}\right)\left(1-\frac{p}{p^*}\right).
\end{aligned} \]
From heuristic reasoning it is clear that for large $z$ the fixation probability $\varphi$ does not change under slight variation of $p$ and $z$, i.e. $\Delta \varphi=\varphi(p_{t+\delta t},z_{t+\delta t})-\varphi(p_t,z_t)$ vanishes with $z_{t+\delta t}=z_t-\delta z$ and 
\[ \begin{aligned}
	p_{t+\delta t} &= p_t+\frac{z_t p_t(1-p_t)}{d}\left(1-\frac{d}{b}\right)\left(1-\frac{p_t}{p^*}\right)\delta t\\
	&= p_t+\frac{z_t p_t(1-p_t)}{d}\left(1-\frac{d}{b}\right)\left(1-\frac{p_t}{p^*}\right)\delta t\frac{\dot{z}_t}{z_t^2\left(-\frac{p_t^2}{a}-\frac{p_t(1-p_t)}{b}-\frac{p_t(1-p_t)}{c}-\frac{(1-p_t)^2}{d}\right)}\\
	&= p_t+\frac{z_t p_t(1-p_t)}{d}\left(1-\frac{d}{b}\right)\left(1-\frac{p_t}{p^*}\right)\delta t \frac{\dot{z}_t}{-\frac{z_t^2}{d}\left(1-p_t\left(2-\frac{d}{c}-\frac{d}{b}\right)+\left(1-\frac{d}{b}\right)\frac{p_t^2}{p^*}\right)}.
\end{aligned} \]
Forming the formal derivative and noting that $\frac{\partial\varphi}{\partial p}\approx 1$ due to the weak selection assumption we have
\[ 0 \approx \frac{d\varphi}{dt} =\frac{\partial \varphi}{\partial p}\frac{dp}{dt} + \frac{\partial \varphi}{\partial z} \frac{dz}{dt} = \frac{(p_{t+\delta t}-p_t)}{\delta t} + \frac{\partial \varphi}{\partial z}\dot{z}. \]
This gives
\[ \frac{\partial\varphi}{\partial z}\sim \frac{p(1-p)\left(1-\frac{d}{b}\right)\left(1-\frac{p}{p^*}\right)}{z\left(1-p\left(2-\frac{d}{c}-\frac{d}{b}\right)+\left(1-\frac{d}{b}\right)\frac{p^2}{p^*}\right)}, \quad \text{for } z\to \infty\]
and therefore $\psi'(z)\sim \frac{1}{z}$ for $z\to\infty$ under weak selection. This shows that 
\begin{equation}\label{eq:limit1} \psi(z)\sim \ln(z) \text{ as } z\to\infty. \end{equation}
Applying this and Lemma \ref{lem:homogeneous} (a) to equation \eqref{eq:psi_new_def} we find for $z\to\infty$ that
\[ \begin{aligned}
	B'(z) = (\psi'(z) + \psi(z)h(z)) e^{-\int_z^\infty h(t)dt} \sim \left(\frac{1}{z} +\frac{\ln(z)}{z^2}\right)e^{\frac{1}{z}} \rightarrow 0, 
\end{aligned} \]
which due to equation \eqref{eq:aux_c} implies that $C(\infty)$ needs to be $0$ in order to provide the right limit behavior of $B(z)$.\\
Lastly, we investigate the limit behavior of $B$ as $z$ tends to $0$. Here, applying equation \eqref{eq:hom_int} we find that equation \eqref{eq:aux_b} for $z\to 0+$ satisfies
\[ B'(z) \sim e^{-2\ln(z^{-2})} 2 \int_z^\infty \frac{Mt+1}{t((\beta+\gamma)d+t)}e^{\ln(t^{-2})} dt \sim z^4 \int_z^\infty \frac{2}{t^3} dt = z^2. \]
Thus, $B$ is integrable at $0+$ giving 
\[ \int_0^z B'(t)dt = B(z)-B(0) = \psi(z)\exp\left(-\int_z^\infty h(t)dt\right)-B(0). \]
The left hand side vanishes for $z\to 0+$ which with equation \eqref{eq:hom_int} implies that 
\[ \psi(z) \sim B(0)\exp\left(\int_z^\infty h(t)dt\right)=\frac{B(0)e^{\alpha}}{z^2}.\]
If $B(0)$ were not $0$ this would mean that $\psi(z)$ is unbounded for $z\to 0+$. But, for very small populations individuals do not sense any density dependence which translates to the competition processes not affecting the fixation probability. Hence, the governing equations read
\[\begin{aligned}
	dx_t &= (\beta-\gamma)x + \frac{1}{\sqrt{M}}\sqrt{(\beta+\gamma)x}dW^1\\
	dy_t &= (\beta-\gamma)y + \frac{1}{\sqrt{M}}\sqrt{(\beta+\gamma)y}dW^2. 
\end{aligned} \]
This system is equivalent to a neutral model, i.e. the fixation probability is independent of the initial population size. Hence, $\psi$ needs to approach $0$ for small $z$ which indeed shows $B(0)=0$. \\
This result and equation \eqref{eq:psi_new_def} allow us to write
\begin{equation}\label{eq:final_psi}
	\begin{aligned}
		\psi(z) &= e^{\int_z^\infty h(t)dt} \int_0^z e^{-2\int_t^\infty h(s)ds+ m(t)} \int_t^\infty \frac{2M + \frac{2}{t}}{(\beta+\gamma)d + t} e^{\int_s^\infty h(u)du -m(t)} ds  dt.
	\end{aligned}
\end{equation}
Thus, we find an explicit form of $\psi$ which indeed shows that it is a non-negative function. This finishes the proof. 
\end{proof}

\begin{remark}
	The positivity of $\psi$ solving equation \eqref{eq:psi_dom} in the case of a dominance game can be shown in i a similar way.
\end{remark}

\section{Numerical evaluation of $\psi(z)$}\label{app:numerics}
\begin{figure*}[ht]
	\subfloat[]{\includegraphics[]{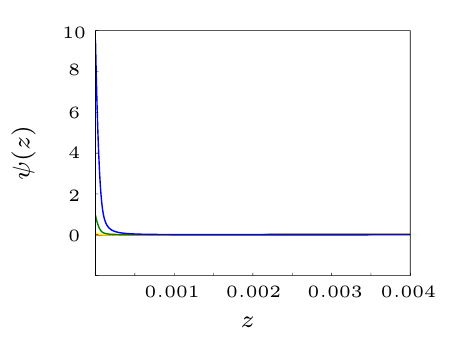}}\quad
	\subfloat[]{\includegraphics[]{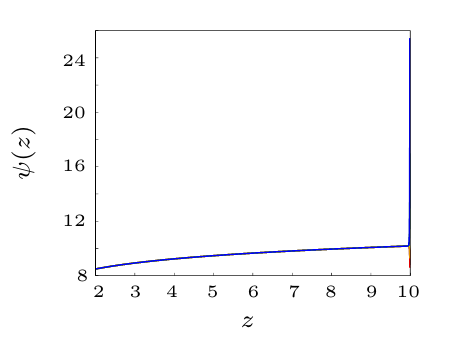}}
	\caption{The function $\psi(z)$ is plotted under varying left boundary conditions (a) and varying right boundary conditions (b).}
	\label{fig:psi}
\end{figure*}

In order to calculate values of $\psi$ we solve equation \eqref{eq:psi} numerically. For this we use the predefined function ``solve\_bvp'' from the scipy.integrate library in Python, \cite{jones:software:2001}. Therefore we need to input boundary values for the algorithm to work with. In particular we evaluate $\psi$ in the interval $[0.01,10]$ with boundary values $\psi(0.01) = \frac{0.01}{3(\beta+\gamma)d}$ and $\psi(10)=\ln(10)$. The justification for choosing $\psi(10)=\ln(10)$ is based on equation \eqref{eq:limit1} above. On the other hand, for very small values of $z$ we use the following reasoning.\\
We need to examine the behavior of equation \eqref{eq:final_psi} for $z\to 0+$. Therefore, we consider the following notation:
\[ \psi(z) = e^{\int_z^\infty h(t)dt} \int_0^z e^{-2\int_t^\infty h(s)ds}H(t) dt,\]
with
\[ H(z) = e^{-2M(2d\beta\ln((\beta+\gamma)d + z)-z)}\int_z^\infty f(t) e^{2M(2d\beta\ln((\beta+\gamma)d+t)-t)} dt,\]
where
\[ f(z) = \left(\frac{2M + \frac{2}{z}}{(\beta+\gamma)d + z}\right)e^{\int_t^\infty h(t)dt}.\]
Next, using equation \eqref{eq:hom_int} we have for $z\to 0+$ that
\[ f(z) \sim \left(\frac{2M + \frac{2}{z}}{(\beta+\gamma)d + z}\right)\frac{1}{z^2} e^{\alpha} \sim \frac{2}{z^3(\beta+\gamma)d} e^{\alpha} \] 
and therefore 
\[ \begin{aligned}
	H(z) &\sim e^{-2M(2d\beta\ln((\beta+\gamma)d + z)-z)}\int_z^\infty \frac{2}{z^3(\beta+\gamma)d} e^{\alpha} e^{2M(2d\beta\ln((\beta+\gamma)d+t)-t)} dt\\
	&\sim \frac{1}{z^2(\beta+\gamma)d}e^{\alpha}.
\end{aligned}\]
Inserting this into $\psi$ we first get
\[ \begin{aligned}
	\int_0^z e^{-2\int_t^\infty h(s)ds}H(t) dt &\sim \int_0^z e^{-2\alpha} t^4 \frac{1}{t^2(\beta+\gamma)d}e^{\alpha}dt\\
	&= \int_0^z e^{-\alpha} \frac{t^2}{(\beta+\gamma)d} dt\\
	&= \frac{z^3}{3(\beta+\gamma)d} e^{-\alpha},
\end{aligned} \]
which yields
\[ \psi(z) \sim e^{\int_z^\infty h(t)dt} \frac{z^3}{3(\beta+\gamma)d} e^{-\alpha} = z^{-2}e^{\alpha}\frac{z^3}{3(\beta+\gamma)d}e^{-\alpha} = \frac{z}{3(\beta+\gamma)d}.\]
This shows the limit behavior of $\psi$ for $z\to 0+$. Note, that the limit behavior for $z\to 0+$ can also be read off equation \eqref{eq:psi} by only considering the leading terms in $z$.\\
Finally, we show in Figure \ref{fig:psi} how the function $\psi$ depends on different choices of boundary values. In subfigure (a) we varied the initial values at $z=0.0001$ from $10^{-8}$ to $10$ while setting $\psi(10)=\ln(10)$. We see that it only affects the values very close to the initial point of the numerical implementation. The same holds if we vary the boundary values at $z=10$, i.e. $\psi(10)$ goes from $1$ to $100$. Holding the initial value $\psi(0.0001)=\frac{0.0001}{3(\beta+\gamma)d}$ fixed we only see variation close to the right boundary. The last subfigure is a plot of $\psi$ with the initial values given by the calculated values above.

\end{appendix}

\end{document}